\documentclass[pdflatex,sn-mathphys-num]{sn-jnl}% Math and Physical Sciences Numbered Reference Style
\usepackage{graphicx}%
\usepackage{multirow}%
\usepackage{amsmath,amssymb,amsfonts}%
\usepackage{amsthm}%
\usepackage{mathrsfs}%
\usepackage[title]{appendix}%
\usepackage{xcolor}%
\usepackage{textcomp}%
\usepackage{manyfoot}%
\usepackage{booktabs}%
\usepackage{listings}%
\usepackage{pgf}
\usepackage{adjustbox}
\usepackage{tikz}
\usepackage{lmodern}
\usepackage[strings]{underscore}
\usepackage{import}

\everymath=\expandafter{\the\everymath\displaystyle}
\usepackage{lmodern}

\newcommand{\mylab}[3]{\raisebox{#2}[0mm][0mm]{%
		\makebox[0mm][l]{\hspace*{#1}\textbf{#3}}}}

\def\aaa{{\it a}}
\def\bbb{{\it b}}
\def\ccc{{\it c}}
\def\ddd{{\it d}}

\theoremstyle{thmstyleone}%
\theoremstyle{thmstyletwo}%

\theoremstyle{thmstylethree}%

\begin{document}

\title[Development of a class of dynamic RANS closures]{Development of a class of dynamic RANS closures for coarse-grained approaches to turbulence}


\author[1]{\fnm{Yves Tessier} \sur{Urrecha}}%\email{yves.tessier@unimore.it}

\author[1]{\fnm{Giorgio} \sur{Manzi}}%\email{259034@studenti.unimore.it}

\author[1]{\fnm{Filippo} \sur{Cancellotti}}%\email{316347@studenti.unimore.it}

\author*[1]{\fnm{Andrea} \sur{Cimarelli}}\email{andrea.cimarelli@unimore.it}

\affil[1]{\orgdiv{Department of engineering "Enzo Ferrari"}, \orgname{University of Modena and Reggio Emilia}, \orgaddress{\street{Via Vivarelli 10}, \city{Modena}, \postcode{41125}, \country{Italy}}}


\abstract{This work represents a crucial advancement in the development of a new class of turbulence closures conceived by \citet{Noi}. This new class of closures is based on the theoretical framework provided by the temporally filtered Navier-Stokes equations. This framework represents a unifying formalism for coarse-grained approaches to turbulence ranging from the statistical to the scale-resolving ones and provides exact equations relating turbulent stresses at different levels of coarsening. The latter have been used in \citet{Noi} as the theoretical foundation to derive a new class of closures based on the dynamic computation of the model coefficient $c_\mu$ appearing in the definition of eddy viscosity. Here, we extend this class of dynamic RANS models by deriving a new formulation based on the $k-\omega$ model including an alternative version where the non-physical decay of free-stream turbulence reproduced by the transport equations of $k$ and $\omega$ is solved in an elegant way. A scale-resolving simulation campaign is performed and the analysis of the results reveals interesting features. Among others, the new dynamic $k-\omega$ model demonstrates strong robustness to a coarsening of the temporal resolution and successfully captures complex flow phenomena, including laminar-to-turbulent transition and the sensitivity of boundary-layer separation and reattachment to free-stream turbulence levels prescribed at the boundaries.}

%%================================%%
%% Sample for structured abstract %%
%%================================%%




\maketitle

%%%%%%%%%%%%%%%%%%%%%%%%%%%%%%%%%%%%%%%%%%%%%%%%%%%%%%%%%%%%%%%%%%%%%%%%%%%%%%%%%%

\section{Introduction}\label{Intro}

In the current state of the art, the dominant approach to CFD in industrial applications stems from the Reynolds-Averaged Navier-Stokes equations (RANS) and from the Large-Eddy Simulation (LES). The first RANS approach aims at describing the ensemble average solution and, hence, it represents a statistical approach to turbulence. On the other hand, the second LES approach aims at describing the large-scale features of the solution and, hence, it represents a scale-resolving approach to turbulence. A third hybrid approach exists, which is being increasingly used in CFD practice. It generally consists of a combination of closures from the RANS and LES realms that enables the computation of flow solutions that simultaneously contain scale-resolving and statistical features depending on the type of flow regions and on the level of discretization employed \citep{menter2005scale, frohlich2008hybrid, spalart2009detached, heinz2020review}. The hybrid nature of these solutions requires the use of an alternative exact theoretical framework able to describe them that goes beyond the classical average and spatially-filtered Navier-Stokes equations. In this context, the temporally filtered Navier-Stokes equations \citep{Pruett2003, Pruett2008} represent the most promising candidate,
\begin{equation}
	\left .
	\begin{split}
		&\frac{\partial \langle u_i \rangle_\tau}{\partial x_i} = 0 \\
		&\frac{\partial \langle u_i \rangle_\tau}{\partial t} + \frac{\partial \langle u_i  \rangle_\tau \langle u_j \rangle_\tau}{\partial x_j} = -\frac{1}{\rho} \frac{\partial \langle p \rangle_\tau}{\partial x_i} + \nu \frac{\partial^2 \langle u_i \rangle_\tau}{\partial x_j\partial x_j} - \frac{\partial R_\tau(u_i,u_j)}{\partial x_j}
	\end{split}
	\right \}
	\label{eq_filter}
\end{equation}
where $\langle \cdot \rangle_\tau$ denoted a temporal filtering operator with characteristic time scale $\tau$. This more general definition of coarse-grained approach to turbulence admits flow solutions ranging from the statistical steady state to the scale resolving \citep{KdF}, depending on the ratio between the filter time scale $\tau$ and the characteristic time scale of turbulence. In other words,  the alternative formalism provided by the time filtering operator offers a more flexible exact theoretical framework which is compatible with hybrid approaches. The potential of this change of paradigm has been addressed in \citet{FadaiGhotbi} where it is shown that the temporal filtering formalism can also be exploited to improve turbulence closures. 

As shown in \citet{Noi}, a notable benefit from operating within the framework of the time filtering formalism is that exact algebraic relations between solutions at different filter levels as those developed for the spatial filtering \citep{Germano1991, Germano1992}, can be leveraged to improve eddy viscosity modelling approaches. A new class of turbulence closures based on the dynamic computation of the model coefficient $c_\mu$ appearing in the eddy viscosity has been conceived in \citet{Noi} and developed for the $k-\varepsilon$ model. The model performances have been tested on the transitional flow over a rounded leading-edge flat plate, on the flow around a rectangular cylinder and on the flow in a turbulent channel with a bump. A remarkable improvement of the accuracy of the solution with respect to the underlying closure is found, especially concerning its ability to capture scale-resolving features and laminar to turbulence transitional phenomena. The aim of the present work is to further assess the potential of the class of dynamic RANS model proposed in \citet{Noi}, by deriving an alternative closure based on the more refined $k-\omega$ model and by performing a thorough analysis of its main scale-resolving features in different flow cases and of its robustness under different temporal resolutions.

The paper is organised as follows. Section \S\ref{sec_Closures} outlines the derivation of the turbulence closures. Section \S\ref{sec_DynamicSettings} lays down some numerical aspects specific to the dynamic procedure. The flow cases that have been used for testing the models are then introduced in section \S\ref{sec_Cases}, together with the details on the numerical configurations adopted. The results for the transitional flow over a flat plate with a rounded leading edge are reported in \S~\ref{subsec_t3lresults}, whereas the wing simulations at high Reynolds numbers are detailed in \S~\ref{subsec_nacaresults}. Concluding remarks are presented in \S~\ref{sec_Conclusions}. Appendices \S~\ref{sec_App_ke_dynamic} and \S~\ref{sec_App_ko_zero_decay} detail the full sets of equations and coefficients defining the new dynamic $k-\varepsilon$ and $k-\omega$ closures respectively.

%%%%%%%%%%%%%%%%%%%%%%%%%%%%%%%%%%%%%%%%%%%%%%%%%%%%%%%%%%%%%%%%%%%%%%%%%%%%%%%%%%

\section{Dynamic RANS closures}\label{sec_Closures}

As shown in \citet{Noi}, the turbulent stresses associated with the implicit temporal filter imposed by a generic closure and those corresponding to an explicit, coarser filter of time scale $T > \tau$ can be formally written as
\begin{equation}
	\begin{split}
		&R_\tau(u_i,u_j) = \langle u_i u_j \rangle_\tau - \langle u_i \rangle_\tau \langle u_j \rangle_\tau \\
		&R_{T_\tau}(u_i,u_j) = \langle \langle u_i u_j \rangle_\tau \rangle_T - \langle \langle u_i \rangle_\tau \rangle_T \langle \langle u_j \rangle_\tau \rangle_T
	\end{split}
	\label{eq_RRSS}
\end{equation}
and can be used to derive a temporal analogue of the Germano's identity \citep{Germano1992},
\begin{equation}
R_{T_\tau}(u_i,u_j) - \langle R_\tau(u_i,u_j) \rangle_T = \langle \langle u_i \rangle_\tau \langle u_j \rangle_\tau \rangle_T - \langle \langle u_i \rangle_\tau \rangle_T \langle \langle u_j \rangle_\tau \rangle_T
\label{eq_Germano}
\end{equation}
The left hand side can take a number of forms depending on the particular model employed for the turbulent stresses, but may invariably be expressed as the product of a model coefficient $c_\mu$ and a tensor $\mathcal{M}_{ij}$ containing only known quantities, i.e. $R_{T_\tau}(u_i,u_j) - \langle R_\tau(u_i,u_j) \rangle_T = c_\mu \, \mathcal{M}_{ij}$. Hence, the identity (\ref{eq_Germano}) can be recast as a system of equations for the model coefficient as
\begin{equation}
	c_\mu \, \mathcal{M}_{ij} = \mathcal{L}_{ij}
	\label{eq_Compact}
\end{equation}
where $ \mathcal{L}_{ij} \equiv \langle \langle u_i \rangle_\tau \langle u_j \rangle_\tau \rangle_T - \langle \langle u_i \rangle_\tau \rangle_T \langle \langle u_j \rangle_\tau \rangle_T$. Thus, a dynamic definition of the model coefficient $c_\mu$ can be derived that minimises the square error of the system of equations (\ref{eq_Compact}) as suggested in \citet{Lilly},
\begin{equation}
	c_\mu = \frac{\mathcal{M}_{ij} \mathcal{L}_{ij}}{\mathcal{M}_{ij} \mathcal{M}_{ij}}
	\label{eq_Lilly}
\end{equation}

\subsection{Dynamic $k-\varepsilon$ model}
\label{subsec_ke}

In this section we apply the temporal filtering formalism and the related dynamic procedure to the well-known $k-\varepsilon$ model \citep{LS1974}, in particular its low Reynolds number formulation by \citet{abe1995new}. The resulting dynamic $k-\varepsilon$ model has been first derived in \citet{Noi} but, for the sake of completeness, we report here its main derivation. Under the $k-\varepsilon$ formulation, the model for the turbulent stresses at either filter levels can be written as
\begin{equation}
	\begin{split}
		&R_\tau(u_i,u_j) = -2 \nu_\tau \langle S_{ij} \rangle_\tau = -2 c_\mu \frac{k_\tau^2}{\varepsilon_\tau} \langle S_{ij} \rangle_\tau \\
		&R_{T_\tau}(u_i,u_j) = -2 \nu_{T_\tau} \langle \langle S_{ij} \rangle_\tau \rangle_T = -2 c_\mu \frac{k_{T_\tau}^2}{\varepsilon_{T_\tau}} \langle \langle S_{ij} \rangle_\tau \rangle_T
	\end{split}
	\label{eq_boussinesq}
\end{equation}
By substituting these expressions into identity (\ref{eq_Germano}), we obtain
\begin{equation}
	2 c_\mu \left \langle \frac{k_\tau^2}{\varepsilon_\tau} \langle S_{ij} \rangle_\tau \right \rangle_T - 2 c_\mu \frac{k_{T_\tau}^2}{\varepsilon_{T_\tau}} \langle \langle S_{ij} \rangle_\tau \rangle_T = \langle \langle u_i \rangle_\tau \langle u_j \rangle_\tau \rangle_T - \langle \langle u_i \rangle_\tau \rangle_T \langle \langle u_j \rangle_\tau \rangle_T
	\label{eq_GermaNoi}
\end{equation}
from which we can define the tensor,
\begin{equation}
	\mathcal{M}_{ij}^{k-\varepsilon} = 2 \left \langle \frac{k_\tau^2}{\varepsilon_\tau} \langle S_{ij} \rangle_\tau \right \rangle_T - 2 \frac{k_{T_\tau}^2}{\varepsilon_{T_\tau}} \langle \langle S_{ij} \rangle_\tau \rangle_T
	\label{eq_M}
\end{equation}
that can be used in equation (\ref{eq_Lilly}) to evaluate the model coefficient for the $k-\varepsilon$ formulation. In the above equations, $k_\tau$ and $\varepsilon_\tau$ are quantities computed through their transport equations,
\begin{equation}
	\left .
	\begin{split}
		&\frac{\partial k_\tau}{\partial t} + \frac{\partial k_\tau \langle u_j \rangle_\tau}{\partial x_j} = 2 \nu_\tau \langle S_{ij} \rangle_\tau \langle S_{ij} \rangle_\tau  - \varepsilon_\tau + \frac{\partial}{\partial x_j} \left[ 
		\left ( \nu + \frac{\nu_\tau}{\sigma^k} \right) \frac{\partial k_\tau}{\partial x_j} \right] \\
		&\frac{\partial \varepsilon_\tau}{\partial t} + \frac{\partial \varepsilon_\tau \langle u_j \rangle_\tau}{\partial x_j} = \frac{\varepsilon_\tau}{k_\tau} \bigg ( C_{\varepsilon_1} 2 \nu_\tau \langle S_{ij} \rangle_\tau \langle S_{ij} \rangle_\tau  - C_{\varepsilon_2} f_\epsilon \varepsilon_\tau \bigg ) + \frac{\partial}{\partial x_j} \left[ 
		\left ( \nu + \frac{\nu_\tau}{\sigma^\varepsilon} \right) \frac{\partial \varepsilon_\tau}{\partial x_j} \right]
	\end{split}
	\right \}
	\label{eq_k_eps_evo}
\end{equation}
where $\sigma^k$, $\sigma^\varepsilon$, $C_{\varepsilon_1}$ and $C_{\varepsilon_2}$ are adjustable model coefficients \citep{Launder} and $f_\epsilon$ is a near-wall damping function to correct the near-wall behaviour \citep{abe1995new}. On the other hand, $k_{T_\tau}$ and $\varepsilon_{T_\tau}$ are defined as
\begin{equation}
	\begin{split}
		&k_{T_\tau} = \langle k_\tau \rangle_T + \mathcal{L}_{ii} / 2\\
		&\varepsilon_{T_\tau} = \langle \varepsilon_\tau \rangle_T + \nu \mathcal{D}_{ii}
	\end{split}
	\label{eq_estimate_k_e}
\end{equation}
where $\mathcal{D}_{ij}$ is a tensor that stems from the same algebraic property of hierarchically organised filters exploited by the double filtering formalism itself,
\begin{equation}
	\begin{split}
		\mathcal{D}_{ij} \equiv &R_{T_\tau} \left (\frac{\partial u_i}{\partial x_k},\frac{\partial u_j}{\partial x_k} \right ) - \left \langle R_\tau\left (\frac{\partial u_i}{\partial x_k},\frac{\partial u_j}{\partial x_k} \right ) \right \rangle_T = \\
		& \left \langle \frac{\partial \langle u_i \rangle_\tau}{\partial x_k} \frac{\partial \langle u_j \rangle_\tau}{\partial x_k} \right \rangle_T - \frac{\partial \langle \langle u_i \rangle_\tau \rangle_T}{\partial x_k} \frac{\partial \langle \langle u_j \rangle_\tau \rangle_T}{\partial x_k}
	\end{split}
\end{equation}
For further details about the derivation of the dynamic $k-\varepsilon$ model, the reader is referred to \citet{Noi}. On the other hand, the full sets of equations and coefficients defining the new dynamic $k-\varepsilon$ are reported in \S~\ref{sec_App_ke_dynamic}. Finally, the implementation of the model for the open-source software OpenFOAM \citep{OF} is available at the following link (\href{https://github.com/GiorgioManzi/DynamicEddyViscosity.git}{github.com/GiorgioManzi/DynamicEddyViscosity.git}).

\subsection{Dynamic $k-\omega$ model}\label{subsec_ko}

In this section, we report, for the first time, a temporally-filtered dynamic formulation of the $k-\omega$ model. The $k-\omega$ model \citep{Wilcox1998} is generally regarded as a more refined approach, due to the more robust behaviour near the walls and in non-equilibrium regions compared with the $k-\varepsilon$ model. On the other hand, the $k-\omega$ model is known to be very sensitive to the free-stream conditions compared with the $k-\varepsilon$ formulation. The classical implementation of the $k-\omega$ model is based on the specific dissipation $\omega_\tau$ that is related with the $k-\varepsilon$ model dissipation as $\omega_\tau = \varepsilon_\tau / (\beta^* k_\tau)$ with $\beta^* = 0.09$. The resulting eddy viscosity reads
\begin{equation}
	\nu_\tau = c_\mu \frac{k_\tau}{\beta^*\omega_\tau}
\end{equation}
By using this formulation as basic turbulence closure, we derive a dynamic procedure for the evaluation of the model coefficient $c_\mu$ that otherwise would be classically prescribed to be constant, $c_\mu = \beta^* = 0.09$.

Within the temporally-filtered formalism, we can define the turbulent stresses at two filter levels that, under the $k-\omega$ formalism, can be written as
\begin{equation}
	\begin{split}
		&R_\tau(u_i,u_j) = -2 \nu_\tau \langle S_{ij} \rangle_\tau = -2 c_{\mu} \frac{k_\tau}{\beta^*\omega_\tau} \langle S_{ij} \rangle_\tau \\
		&R_{T_\tau}(u_i,u_j) = -2 \nu_{T_\tau} \langle \langle S_{ij} \rangle_\tau \rangle_T = -2 c_{\mu} \frac{k_{T_\tau}}{\beta^*\omega_{T_\tau}} \langle \langle S_{ij} \rangle_\tau \rangle_T
	\end{split}
	\label{eq_boussinesq_k_omega}
\end{equation}
that can be substituted into identity (\ref{eq_Germano}) thus leading to the following system of equations,
\begin{equation}
	2 c_{\mu} \left\langle \frac{k_{\tau}}{\beta^*\omega_{\tau}} \langle S_{ij} \rangle_{\tau} \right\rangle_{T} 
	- 2 c_{\mu} \frac{k_{T_\tau}}{\beta^*\omega_{T_\tau}} \langle \langle S_{ij} \rangle_{\tau} \rangle_{T} 
	= \langle \langle u_i \rangle_{\tau} \langle u_j \rangle_{\tau} \rangle_{T} 
	- \langle \langle u_i \rangle_{\tau} \rangle_{T} \langle \langle u_j \rangle_{\tau} \rangle_{T}
\end{equation}
from which we can define the tensor,
\begin{equation}
\mathcal{M}_{ij}^{k-\omega} = 2 \left\langle \frac{k{\tau}}{\beta^*\omega_{\tau}} \langle S_{ij} \rangle_{\tau} \right\rangle_{T} 
- 2 \frac{k_{T_\tau}}{\beta^*\omega_{T_\tau}} \langle \langle S_{ij} \rangle_{\tau} \rangle_{T}
\end{equation}
that can be used in equation (\ref{eq_Lilly}) to dynamically compute the model coefficient $c_\mu$. In the above equations, the turbulent quantities $k_{\tau}$ and $\omega_{\tau}$ are computed through their transport equations,
\begin{equation}
	\left .
	\begin{split}
		&\frac{\partial k_{\tau}}{\partial t} + \frac{\partial k_{\tau}\langle u_j \rangle_{\tau}}{\partial x_j}
		= 2 \nu_{\tau} \langle S_{ij} \rangle_{\tau} \langle S_{ij} \rangle_{\tau} - \beta^* k_{\tau} \omega_{\tau}
		+ \frac{\partial}{\partial x_j} \left[ \left( \nu + \frac{\nu_{\tau}}{\sigma^k} \right) \frac{\partial k_{\tau}}{\partial x_j} \right] \\
		&\frac{\partial \omega_{\tau}}{\partial t} + \frac{\partial \omega_{\tau}\langle u_j \rangle_{\tau}}{\partial x_j}
		= \frac{\omega_{\tau}}{k_{\tau}} \left( 2 \gamma \, \nu_\tau \langle S_{ij} \rangle_\tau \langle S_{ij} \rangle_\tau - \beta k_{\tau} \omega_{\tau} \right)
		+ \frac{\partial}{\partial x_j} \left[ \left( \nu + \frac{\nu_{\tau}}{\sigma^\omega} \right) \frac{\partial \omega_{\tau}}{\partial x_j} \right]
	\end{split}
	\right \}
	\label{eq_k_omega_ivo}
\end{equation}
where $\sigma^k$, $\sigma^\omega$, $\gamma$ and $\beta$ are adjustable model coefficients \citep{Wilcox1988}. On the other hand, the turbulent quantities at the test filter level are evaluated as
\begin{equation}
	\begin{split}
		&k_{T_\tau} = \langle k_\tau \rangle_T + \mathcal{L}_{ii} / 2\\
		&\omega_{T_\tau} = \frac{\varepsilon_{T_\tau}}{\beta^*k_{T_\tau}} = \frac{1}{k_{T_\tau}} \left( \left\langle k_{\tau} \omega_{\tau} \right\rangle_{T} + \frac{\nu}{\beta^*} \mathcal{D}_{ii} \right)
	\end{split}
	\label{eq_estimate_k_w}
\end{equation}
where $k_{T_{\tau}}$ maintains the exact same form used in equation (\ref{eq_estimate_k_e}) for the dynamic $k-\varepsilon$ model, while the expression for the specific dissipation rate $\omega_{T_{\tau}}$ is straightforwardly developed from that of dissipation $\varepsilon_{T_{\tau}}$ again reported in equation (\ref{eq_estimate_k_e}).

A complete view of the full sets of equations and coefficients defining the new dynamic $k-\omega$ is reported in \S~\ref{sec_App_ko_zero_decay}. Furthermore, the implementation of the model for the open-source software OpenFOAM \citep{OF} is available at the following link (\href{https://github.com/GiorgioManzi/DynamicEddyViscosity.git}{github.com/GiorgioManzi/DynamicEddyViscosity.git}).

\subsection{Zero-decay dynamic $k-\omega$ model}\label{subsec_zerodecay}

In this section we propose an alternative version of the dynamic $k-\omega$ model that is based on arguments related to the transport equations (\ref{eq_k_omega_ivo}) of the turbulent quantities $k_{\tau}$ and $\omega_{\tau}$. Indeed, the presence of $c_\mu$ in the production of both $k_{\tau}$ and $\omega_{\tau}$ can alter the global equilibrium on which the broader source term is founded \citep{Launder, Wilcox1988, Pope, Malalasekera}. This is because the variation of $c_\mu$ away from its conventional value changes the magnitude of the production terms in a manner that is not mirrored by the destruction terms, where $c_\mu$ does not explicitly appear. Therefore, if consistency with the global equilibrium assumption is to be maintained, a commensurate adjustment to the destruction term is warranted, which involves multiplying the sink terms of equations (\ref{eq_k_omega_ivo}) by the instantaneous value of the dynamic coefficient $c_\mu$ normalised with respect to its conventional value $c_{\mu,0}=0.09$,
\begin{equation}
	\left .
	\begin{split}
		&\frac{\partial k_{\tau}}{\partial t} + \frac{\partial k_{\tau}\langle u_j \rangle_{\tau}}{\partial x_j}
		= 2 \nu_{\tau} \langle S_{ij} \rangle_{\tau} \langle S_{ij} \rangle_{\tau} - c_{\mu}^*\beta^* k_{\tau} \omega_{\tau}
		+ \frac{\partial}{\partial x_j} \left[ \left( \nu + \frac{\nu_{\tau}}{\sigma^k} \right) \frac{\partial k_{\tau}}{\partial x_j} \right] \\
		&\frac{\partial \omega_{\tau}}{\partial t} + \frac{\partial \omega_{\tau}\langle u_j \rangle_{\tau}}{\partial x_j}
		= \frac{\omega_{\tau}}{k_{\tau}} \left( 2 \gamma \, \nu_\tau \langle S_{ij} \rangle_\tau \langle S_{ij} \rangle_\tau - c_{\mu}^* \beta k_{\tau} \omega_{\tau} \right)
		+ \frac{\partial}{\partial x_j} \left[ \left( \nu + \frac{\nu_{\tau}}{\sigma^\omega} \right) \frac{\partial \omega_{\tau}}{\partial x_j} \right]
	\end{split}
	\right \}
	\label{eq_k_omega_ivo2}
\end{equation}
where $c_{\mu}^* = c_{\mu}/c_{\mu,0}$. This modified approach immediately restores the equilibrium prescribed by the classical $k-\omega$ closure. To note that this is not the sole possible correction of the source terms of the transported quantities. Investigating whether other correction methods outperform the present one, goes beyond the main purpose of the present work and are left to future investigations.

This correction in the transport equations for $k_{\tau}$ and $\omega_{\tau}$ has significant repercussions that go beyond its primary purpose, with remarkable benefits. It is well known that free stream turbulence can be mimicked through the free stream values of the turbulent quantities when using RANS eddy viscosity models, but a precise control over the modelled turbulence is challenging due to the decay of the transported variables in the unperturbed region of the flow. As a matter of fact, it is common practice to design the inlet values for the turbulent quantities to achieve the desired free-stream turbulence conditions at the streamwise location of interest, by compensating for their decay according to the corresponding laws. The process is further complicated by the additional effect of numerical dissipation sensitive not only to discretisation but to the characteristics of the modelled turbulence as well. In the particular case of the $k-\omega$ formulation, the laws of decay read
\begin{equation}
	\left.
	\begin{aligned}
		k_\tau(x) &= k_{in} \left[1 + \beta\, \omega_{in} \frac{x}{U_{0}} \right]^{-\frac{\beta^*}{\beta}} \\
		\omega_\tau(x) &= \frac{\omega_{in}}{1 + \beta\, \omega_{in} \frac{x}{U_{0}}}
	\end{aligned}
	\right\}
\end{equation}
where $k_{in}$ and $\omega_{in}$ are the imposed inlet values for $k_{\tau}$ and $\omega_{\tau}$, and $x$ is the distance from the boundary in the direction of the free stream, whose velocity is $U_0$. These laws directly derive from the transport equations of $k_{\tau}$ and $\omega_{\tau}$ in the unperturbed conditions in the free stream, which reduce to a simple balance between streamwise convection and dissipation. 

In this context, the adoption of the factor $c_{\mu}^*$ for the destruction terms of the transport equations completely changes the scenario. Indeed, the nature of the dynamic procedure is such that the coefficient $c_\mu$, and consequently $c_\mu^*$, becomes null in the regions where turbulent activity does not manifest through temporal oscillations of the velocity field, such as in the unperturbed flow regions of the free stream. As a consequence, the transport equations (\ref{eq_k_omega_ivo2}) reduce to a pure convection equation where inlet boundary conditions are rigidly transported through the unperturbed free-stream regions. In other words, the turbulent quantities $k_{\tau}$ and $\omega_{\tau}$ do not decay in the unperturbed regions of the free stream.

While the decay is well founded in the theory of freely decaying homogeneous isotropic turbulence, there is also a notion with significant support \citep{SpalartDecay} that, indeed, it ought not apply to the free stream region of many flow cases. The most evident example is that of external aerodynamics problems, where free-stream turbulence represents an ambient condition rather than an evolving flow phenomenon: in this context, the decay of homogeneous isotropic turbulence is no longer an adequate proxy for the free-stream evolution of turbulence.

As a matter of fact, a number of solutions have been proposed to circumvent the issues caused by the decay of the turbulent quantities in the free stream, such as the use of "floor values" that the turbulent quantities are not allowed to fall below (as implemented in the NASA PAB3D code \citep{NASAPAB3D} and OpenFOAM), or the ad-hoc prescription of flow regions where the destruction term is suppressed altogether (which is employed in the CFD++ code \citep{CFD++}). With the developed dynamic procedure and the proposed modification of the transport equations (\ref{eq_k_omega_ivo2}), however, we have a completely autonomous solution for preventing the decay of the transported quantities which is merely a consequence of the closure's formulation. Moreover, it holds a distinct advantage for long-body simulations, where matching the desired free stream turbulence characteristics at the body's leading edge does not guarantee their correct reproduction further downstream.

In closing this section, let us point out that the proposed zero-decay modification of the transport equations is sufficiently general to be applicable also to other implementations of the dynamic procedure such as that developed in \citet{Noi} for the $k-\varepsilon$ model and reported in the present work in the previous section \S\ref{subsec_ke}.

\subsection{Control of the unsteadiness}

The developed class of closures should be considered as a hybrid approach in a broad sense, not RANS/LES, but rather statistical/scale-resolving. Strictly in terms of modelling, the only closure it explicitly involves is the underlying RANS model. Hence, the dynamic procedure does not switch between any RANS/LES closures by comparing scales like in more classical hybrid approaches. In particular, the formulation does not include an explicit dependence on any adjustable parameters like the mesh size, the time step or distance from the wall, which means there is no direct control over the transition between the statistical and scale-resolving behaviour. Still, the dependence on the explicitly prescribed spatial and temporal resolution is implicitly recovered by the dynamic procedure itself. In particular, the dynamic RANS procedure is based on a measure of the change of the solution among two time scales through the use of a test filter that is proportional to the time step of the simulation. The spatial resolution determines the type of turbulence structures that can be generated and that span these two time scales. Eventually, the dynamic procedure adapts the eddy viscosity to not dissipate them. In other words, the spatial and temporal filtering implicitly performed by the dynamic eddy viscosity is largely dependent on the numerical resolution adopted, although not explicitly taken into account in the model definition. Accordingly, our approach is found to be sensitive --albeit in a robust manner-- to the grid size as demonstrated in \citet{Noi} and to the time step as shown in section \S\ref{subsubsec_t3lCFL} of the present work.

Overall, we argue that the lack of an explicit management of the statistical-to-scale-resolving transition is a remarkable property of the dynamic RANS formulation, inasmuch as the dynamic procedure is capable of optimizing this transition itself, thereby maximizing the amount of information provided by the solution that are supported by the spatial and temporal discretization. These latter have to be understood in a broad sense, i.e. the actual length and time scales that can be reproduced in the solution that are given by the combination of the grid size and time step with the numerical schemes adopted.

%%%%%%%%%%%%%%%%%%%%%%%%%%%%%%%%%%%%%%%%%%%%%%%%%%%%%%%%%%%%%%%%%%%%%%%%%%%%%%%%%%

\section{Dynamic procedure specific settings}\label{sec_DynamicSettings}

In this section, we provide some specific settings that are demanded by the use of the dynamic procedure for both the dynamic $k-\varepsilon$ and $k-\omega$ models.

\subsection{Test filter}

In order to establish the dynamic definition for $c_\mu$, it is essential to define an appropriate kernel for the explicit test filter and a proper time scale $T$. As shown in \citet{Noi}, the most straightforward option for its kernel to pursue scale-resolving unsteady solutions is the backward moving average,
\begin{equation}
	\langle \phi \rangle_T (N, t) = \frac{1}{N \Delta t} \int_{t-N \Delta t}^{t} \phi(t') dt'
\end{equation}
where $N$ is the number of time steps that enter the moving average, yielding $T=N \Delta t$. The simulations in this study were all executed by using two time steps $N=2$, but it is worth pointing out that a range of values for $T=N \Delta t$ was evaluated, specifically, $N=3$, 5 and 10. The choice of $N=2$ came down to the fact that the alternatives yielded quite similar results but a lower $N$ translates into a lower memory usage.

\subsection{Clipping procedure}\label{subsec_other}

On average, the dynamic $c_{\mu}$ tends to settle on fairly low values across the domain, maxing out around the standard $c_{\mu,0}$ in the most active areas of shear layers and in poorly spatially resolved regions. However, some outlier spikes do occasionally appear in instantaneous solutions, which have been observed to compromise numerical stability. On the other hand, analogously to the LES application \citep{Katz2000}, the dynamic procedure can --and often does-- return negative values, but the potential benefits of a negative eddy viscosity cannot be exploited because it would expel the solution out of the stable region for time integration. In order not to upset the numerical stability of the solution, we adopt a clipping operation to bound the dynamic coefficient to the range $0 \leq c_{\mu} \leq 0.2$. A number of alternatives were evaluated for the upper limit, but the solution is found to be fairly insensitive to it. A value of 0.2 was therefore selected to keep the range roughly centered about the standard $c_{\mu,0}$.

\subsection{$k-\varepsilon$ specific adjustments}\label{subsec_keadjust}

Due to its wall behaviour characteristics, the classical $k-\varepsilon$ formulation requires modification to achieve asymptotic consistency with the low-Re approach \citep{lam1981modified, Chien, AKN}. In particular, a damping function $f_\epsilon$ to adjust the $\varepsilon$ destruction term is indispensable if the equations are to be integrated down to the wall. Conversely, the damping function $f_\mu$ that modulates the eddy viscosity directly must be excluded to allow the dynamic procedure to be fully functional. This function is used to correct the strong tendency of the $k-\varepsilon$ model to overestimate the eddy viscosity near walls, but the dynamic procedure is capable of correctly adapting $\nu_\tau$ to the different flow regions on its own \citep{Noi}, which renders the damping function redundant and counter-productive.

%%%%%%%%%%%%%%%%%%%%%%%%%%%%%%%%%%%%%%%%%%%%%%%%%%%%%%%%%%%%%%%%%%%%%%%%%%%%%%%%%%

\section{Flow cases and numerical configuration}\label{sec_Cases}

In the present work, we study the performance of the new dynamic RANS models in the flow settings of the transitional flow over a rounded leading edge flat plate and of the near-stall flow over a NACA 4412 wing profile. Although the cases are very different, a number of aspects about their numerical configurations are the same. All simulations were run on the open-source code OpenFOAM, and particularly with its solver employing the PISO algorithm for pressure-velocity decoupling. Pressure was handled with the GAMG solver due to the strong parallelisation of the domain, while the transported quantities were addressed with a smooth solver. Upwind-biased 2nd order blended schemes were used for all convection terms, whereas the diffusive terms were calculated with linear interpolation and treated with a limited non-orthogonal correction, and all gradients were discretised with Gaussian integration and bounded. Time integration was performed with the backward Euler scheme, with first-order accuracy for the stresses and second-order for the non-linear terms. Unless otherwise stated, a constant time step was tuned so as to satisfy the condition $CFL < 1.5$. Other features common to both cases are the use of a C-type grid with a structured layout, as well as the prescription of periodic conditions along the statistically homogeneous spanwise direction.

\subsection{T3L flow case}\label{subsec_t3l}

The first case considered in this work is the transitional flow over a rounded leading edge flat plate at $Re = U_0 D/\nu = 3333$ (where $U_0$ is the free-stream velocity, $D$ the plate thickness and $\nu$ the kinematic viscosity) under varying free stream turbulence conditions. This flow configuration, known as the ERCOFTAC T3L, is an interesting benchmark for the present solution because it focuses on the laminar separation, transition, and reattachment of the flow, and how they change with the characteristics of the free stream turbulence, all of which are points of high interest for the approach. The study will address how the dynamic RANS models implemented on OpenFOAM perform in comparison to a high-fidelity implicit LES simulation \citep{Bassi2020}, to a previous dynamic $k-\varepsilon$ implementation developed on a different finite volume code T-Flows (\href{https://github.com/DelNov/T-Flows}{https://github.com/DelNov/T-Flows}), and to state-of-the-art statistical and scale-resolving approaches available on OpenFOAM. Among others, the robustness to increasingly coarse temporal resolution will be examined.

\begin{figure}
	\centering
	\includegraphics[width=0.45\linewidth, trim={0px 0px 0px 0px},clip]{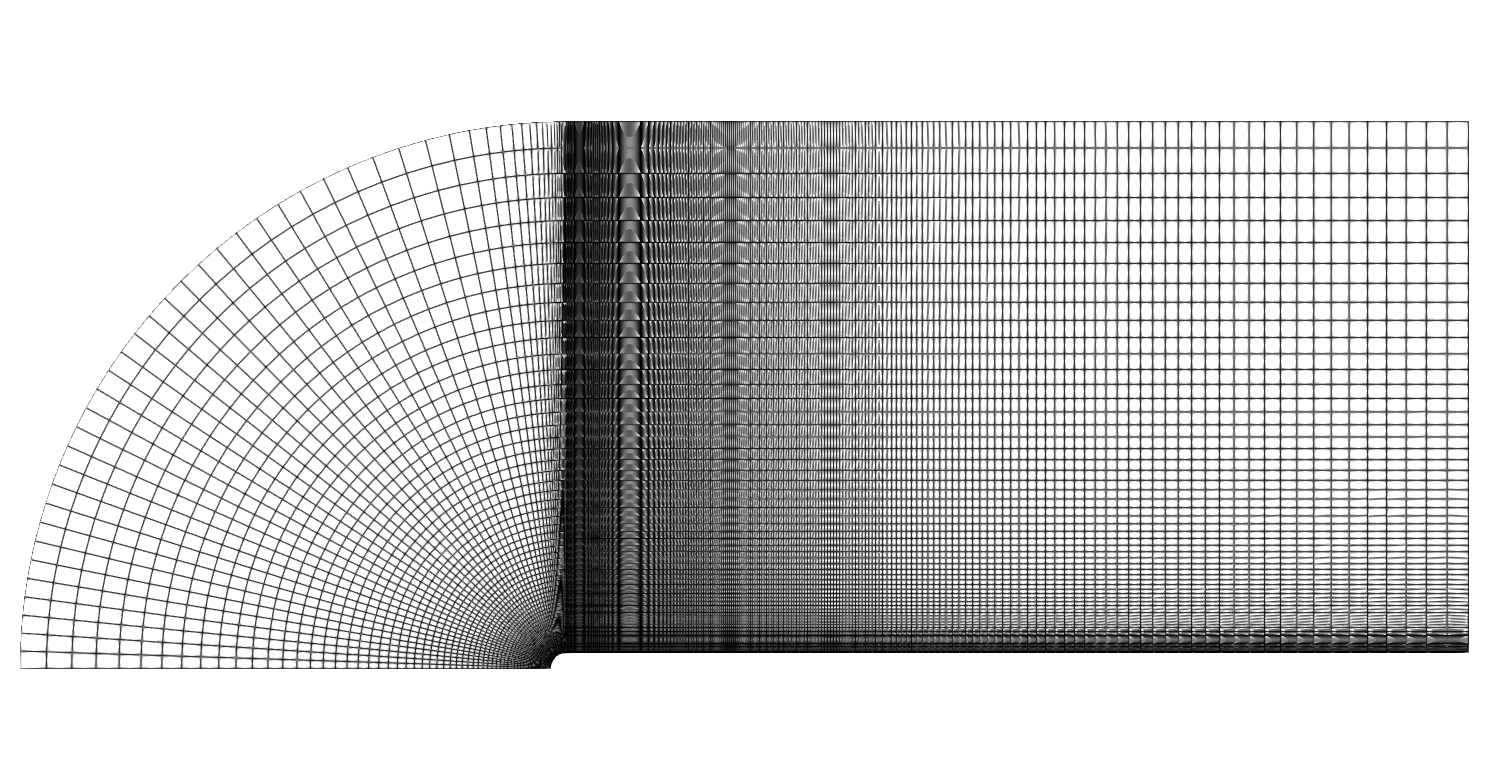}
	\includegraphics[width=0.45\linewidth, trim={0px 0px 0px 0px},clip]{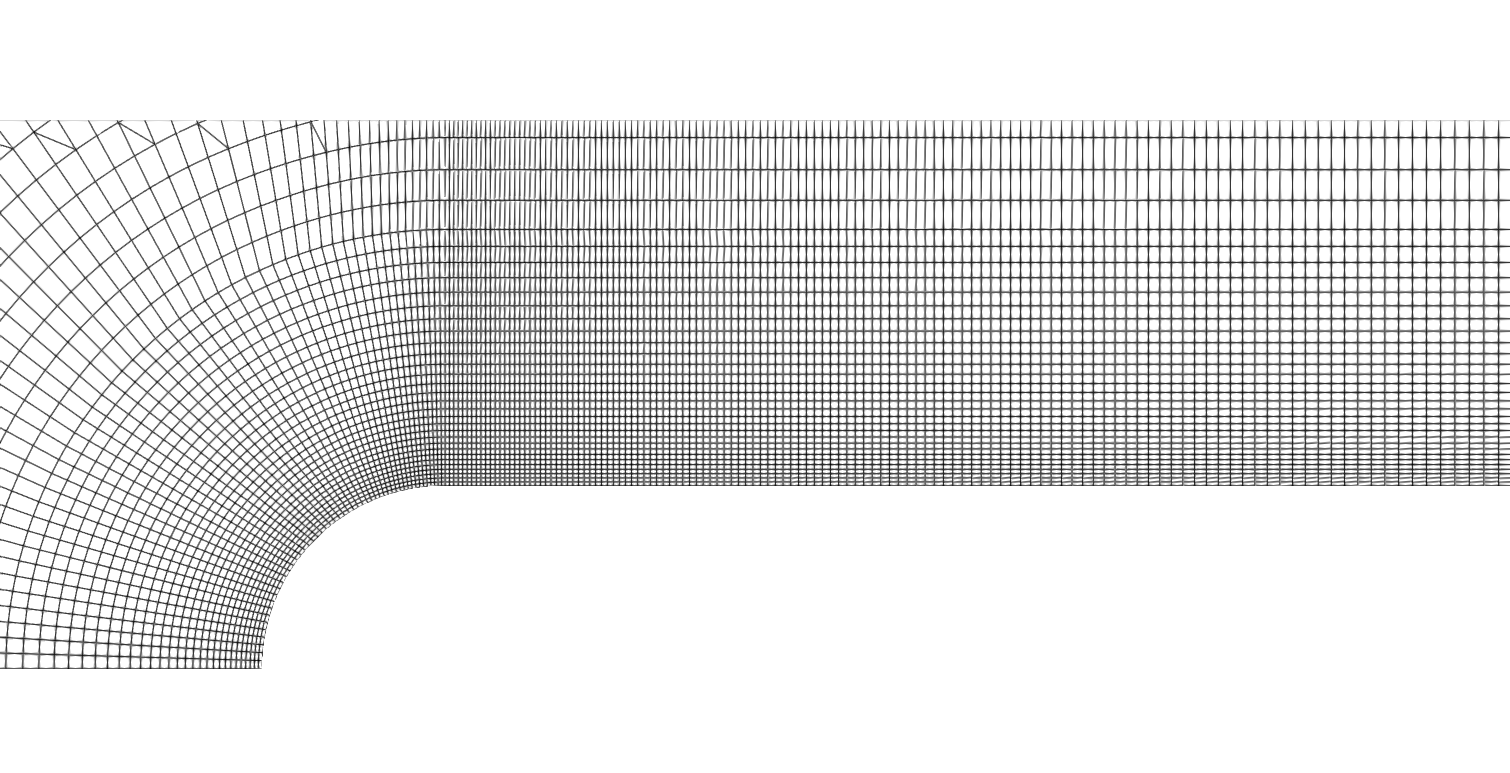}
	\caption{Computational grid for the T3L simulations. The detailed view on the right panel illustrates the refined resolution around the main area of interest [$-1 \leq x/D \leq 4$, $0 \leq y/D \leq 1.5$].}
	\label{fig:Grid_T3L}
\end{figure}

All simulations of the T3L have been performed on the computational grid developed for the simulations reported in \citet{Noi}, which represents only the top half of the domain by virtue of the statistical symmetries of the flow solution, see figure \ref{fig:Grid_T3L}. A conventional fixed velocity inlet and pressure outlet with Neumann conditions are used, whereas free-slip conditions apply at the top of the domain and at the symmetry plane. Periodicity is configured along the statistically homogeneous spanwise direction, while a no-slip condition is applied at the wall of the flat plate.

\begin{figure}
	\centering
	\includegraphics[width=0.45\linewidth, trim={0px 0px 0px 0px},clip]{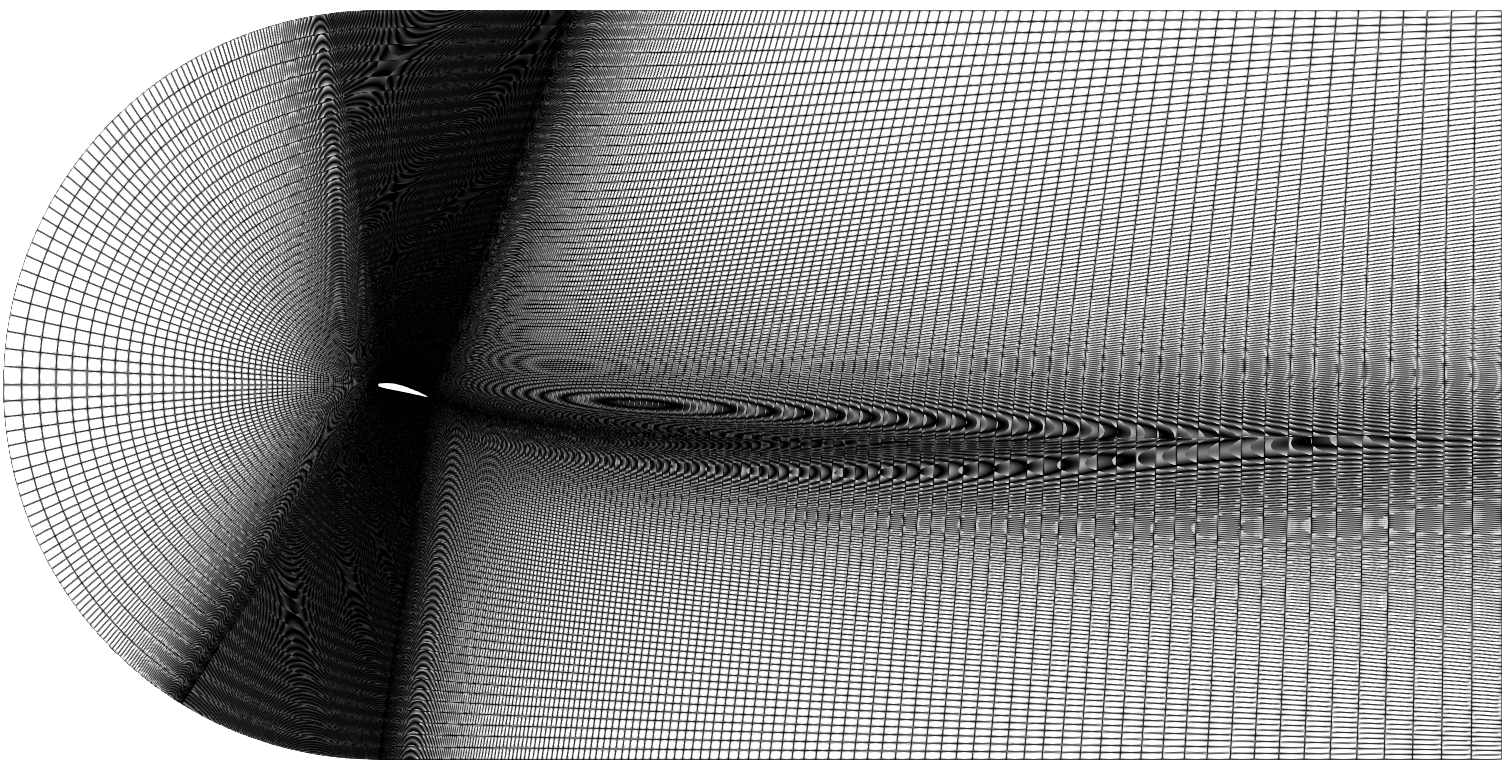}
	\includegraphics[width=0.45\linewidth, trim={0px 0px 0px 0px},clip]{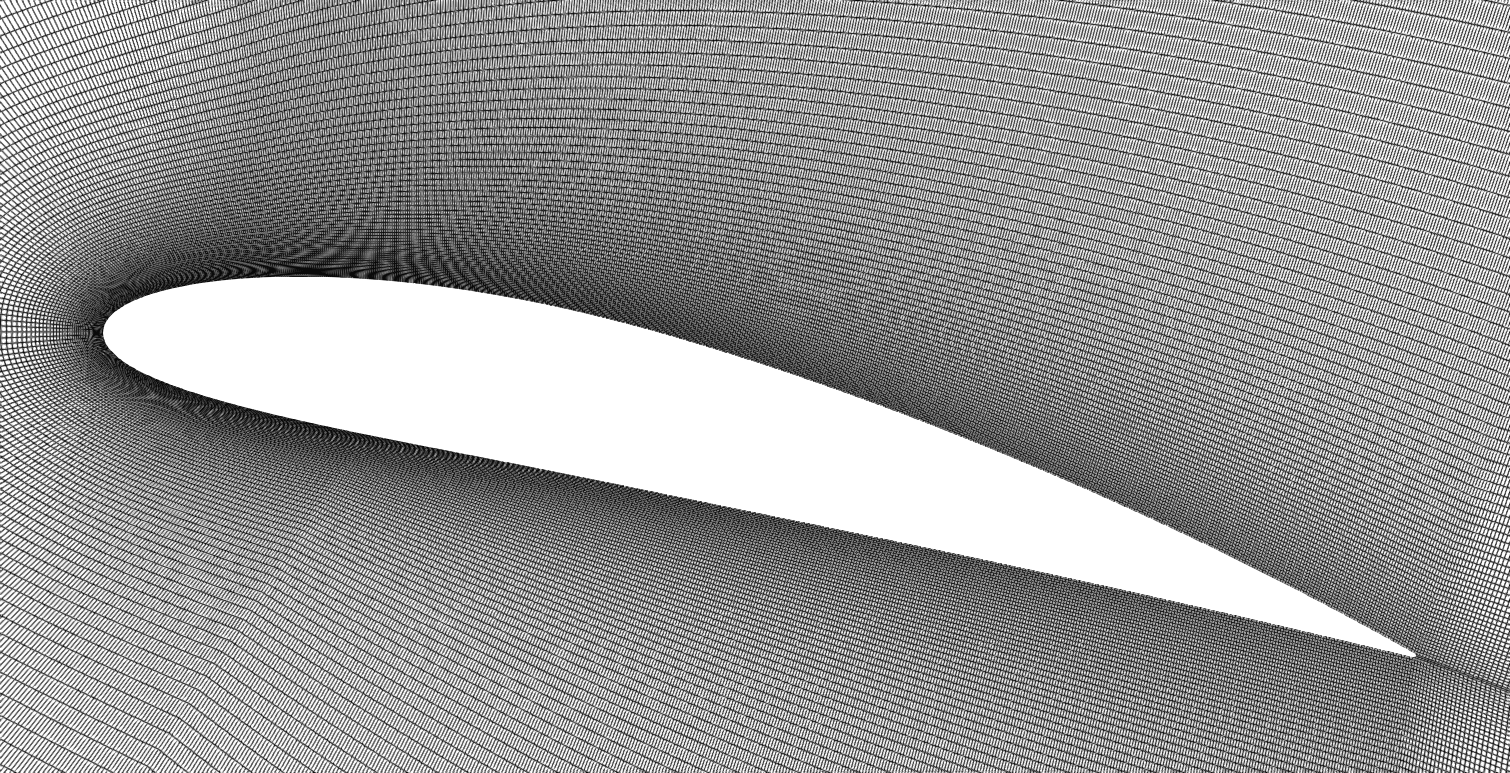}
	\caption{Computational grid for the NACA 4412 simulations. The detailed view on the right panel illustrates the refined resolution around the main area of interest [$-1 \leq x/D \leq 4$, $0 \leq y/D \leq 1.5$].}
	\label{fig:Grid_NACA}
\end{figure}

The domain is bounded in the $x$-streamwise, $y$-vertical, and $z$-spanwise directions between $-17 \leq x/D \leq 28$, $0 \leq y/D \leq 17$, and $0 \leq z/D \leq 2$ respectively, sufficient to accommodate and not restrict the turbulent structures characteristic of the solution. Accordingly, (u, v, w) will hereafter denote the streamwise, vertical and spanwise components of the velocity vector. Naturally, cells are clustered towards the wall and the leading edge of the plate to guarantee a good resolution of the boundary layer, to better capture the main phenomena of the flow and pinpoint the reattachment position. As a result, the area between $0.5 \leq x/D \leq 9.3$ and $0.5 \leq y/D \leq 1.2$ --just above the start of the flat portion of the plate-- is as finely resolved as $\Delta x/D = 0.012 \div 0.1$, $\Delta y/D = 0.01 \div 0.046$ and $\Delta z/D = 0.05$, for a final count of 1.03M cells. Because a low-Re approach is followed, the wall-normal resolution is on average smaller than one friction length thus allowing us to avoid the use of wall functions.

\subsection{NACA 4412 flow case}\label{subsec_naca}

The second case addressed in this work is the flow over a NACA 4412 wing section at near-stall conditions for $Re= U_0 c / \nu= 1.52 \cdot 10^6$ and angle of attack $\alpha=13.87^\circ$ where $c$ is the chord length. This case is selected to evaluate the model's performance in high-Reynolds number regimes solved by using wall functions. This choice aligns with established industrial standards for high-Reynolds number flows, where resolving the entire viscous sublayer is often computationally prohibitive and less common in practical engineering applications. In addition, the selection of this particular configuration is also motivated by the availability of experimental datasets, providing valuable reference data for validation purposes. 

As shown in figure \ref{fig:Grid_NACA}, the computational domain extends for $-7 \leq x/c \leq 18$, $-7 \leq y/c \leq 7$, and $0 \leq z/c \leq 0.6$, where $x$, $y$, and $z$ denote the streamwise, vertical, and spanwise directions, respectively. The grid is composed by 9.02M computational elements that are clustered around the wing section and its wake in order to guarantee the resolution of the relevant phenomena. The resulting spatial resolution adopted near the wall is $\Delta x/c = 0.00135 \div 0.0022$, $\Delta y/c = 0.0005 \div 0.001$ and $\Delta z/c = 0.005$. The wall-normal resolution over the suction side of the wing is on average of the order of 50 friction lengths. As already mentioned, this value has been chosen to study the performance of the dynamic RANS models when employed to study high-Reynolds number flows with wall functions. Finally, a Dirichlet boundary condition is prescribed for the velocity at the C-boundary of the domain. A standard Neumann condition is prescribed at the outlet while periodic boundary conditions are used in the spanwise direction. Boundary layer tripping is not explicitly prescribed in the computational domain; instead, a natural transition approach is favored to ensure a consistent framework for evaluating the dynamic model against prevailing industrial methodologies.

%%%%%%%%%%%%%%%%%%%%%%%%%%%%%%%%%%%%%%%%%%%%%%%%%%%%%%%%%%%%%%%%%%%%%%%%%%%%%%%%%%%%%

\section{Transitional flow over a rounded leading edge flat plate}\label{subsec_t3lresults}

Other than with the dynamic closures here presented, the transitional flow about a rounded leading edge flat plate has also been simulated with the standard $k-\omega$ model, to allow for a direct evaluation of the enhancement attributable to the dynamic procedure. The $k-\omega$ SST formulation \citep{SST} augmented with the DES extension \citep{SSTDES} is also used, to assess the relative performance of a hybrid scale-resolving approach \citep{VonTerzi, Spalart2021} under the same numerical configuration. Results will be compared against reference data from high-fidelity implicit-LES simulations performed by \citet{Bassi2020}. The inlet boundary conditions for the turbulent quantities $k_\tau$ and $\omega_\tau$ will be varied to address three different free-stream turbulence conditions corresponding to three different values of turbulence intensity evaluated at the leading edge as $I = \sqrt{2 k_\tau/3}/U_0$. We consider $I = 0.2\%$, $2.3\%$ and $5.6\%$ that will be hereafter referred to as the cases at low, medium and high free-stream turbulence intensity, respectively. Statistics are extracted by averaging in the statistically homogeneous spanwise direction as well as in time, using 60 samples recorded over $354L/U_0$ times once the solution has reached statistically a fully developed condition. The resulting averaged operator will be hereafter denoted as $\langle \cdot \rangle$.

\subsection{Assessment of the effects of the baseline closure}\label{subsubsec_t3ldyndyndyn}

As introduced in \citet{Noi} and unequivocally demonstrated in section \S\ref{sec_Closures}, the temporal filtering formalism and the related dynamic procedure can be applied to different eddy-viscosity models as baseline closures. It is then relevant to address how the quality of the solution changes by varying the baseline closures. To this aim, results from the dynamic $k-\varepsilon$, dynamic $k-\omega$ and dynamic zero-decay $k-\omega$ will be analysed in this section. The effect of the finite volume solver of the Navier-Stokes equations will be also considered by taking into account results from the dynamic $k-\varepsilon$ obtained by using the T-Flows finite volume solver.

\begin{figure}
	\centering
	\includegraphics[width=0.45\linewidth, trim={0px 0px 0px 0px},clip]{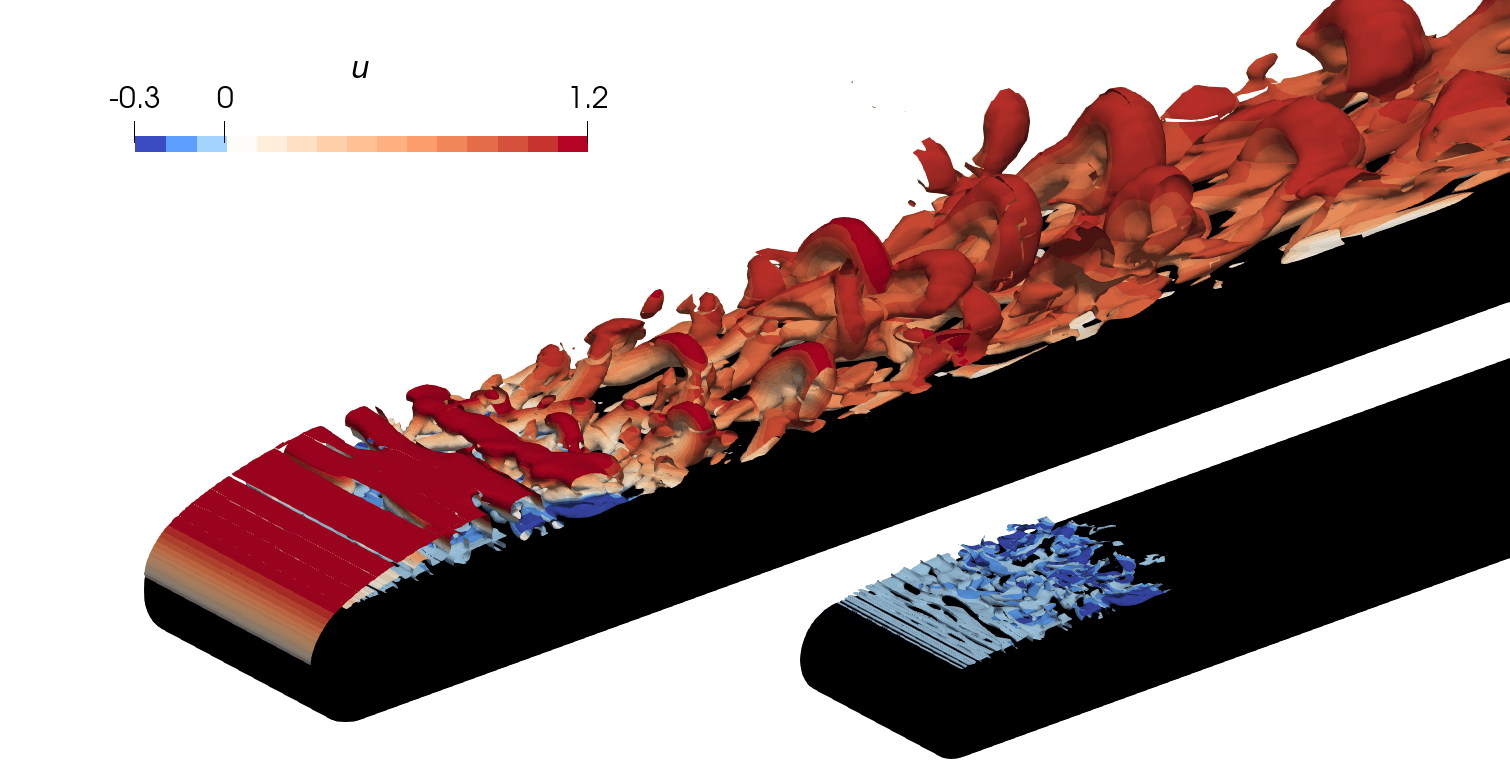}\mylab{-67mm}{17.25mm}{(\aaa)}
	\includegraphics[width=0.45\linewidth, trim={0px 0px 0px 0px},clip]{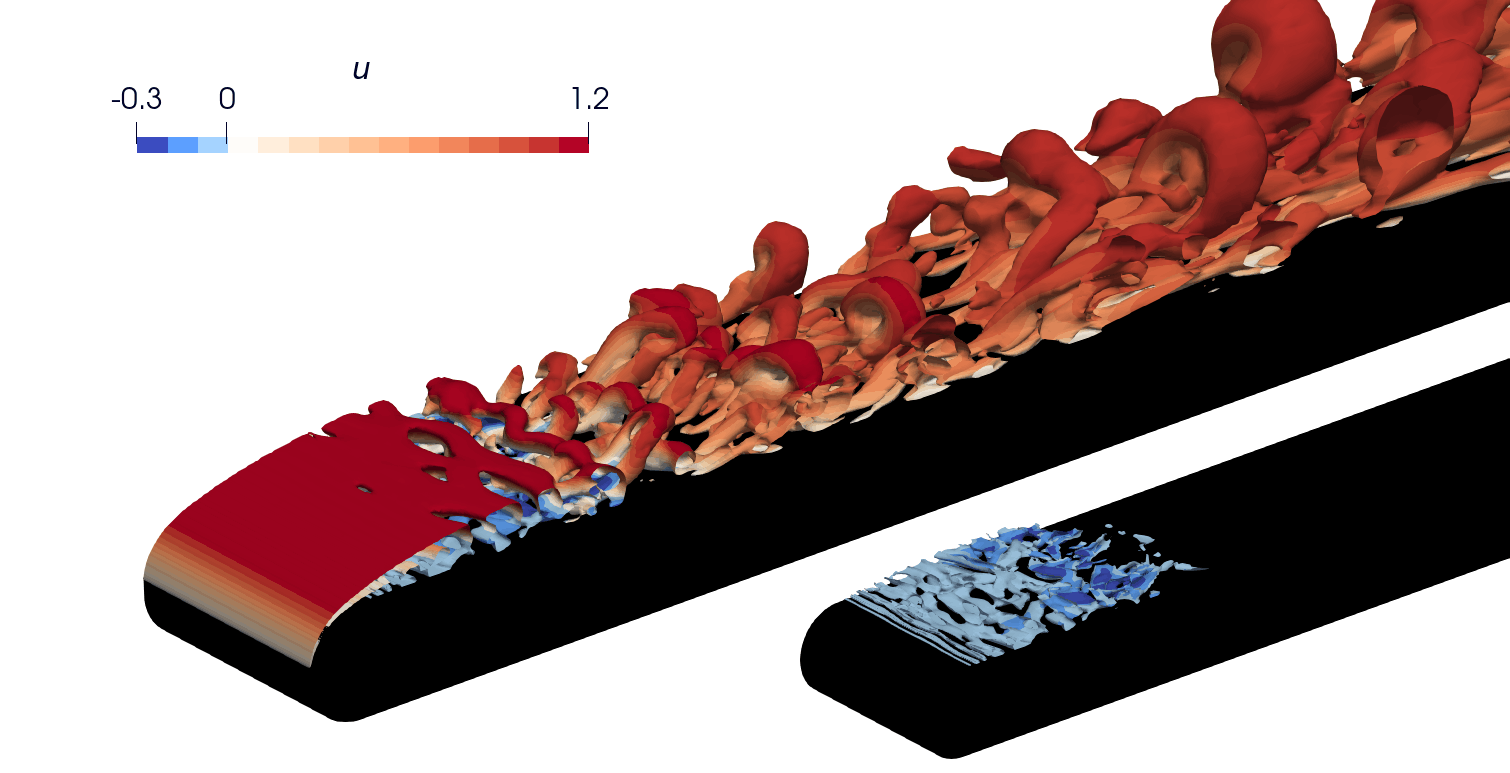}\mylab{0mm}{17.25mm}{(\bbb)}
	
	\includegraphics[width=0.45\linewidth, trim={0px 0px 0px 0px},clip]{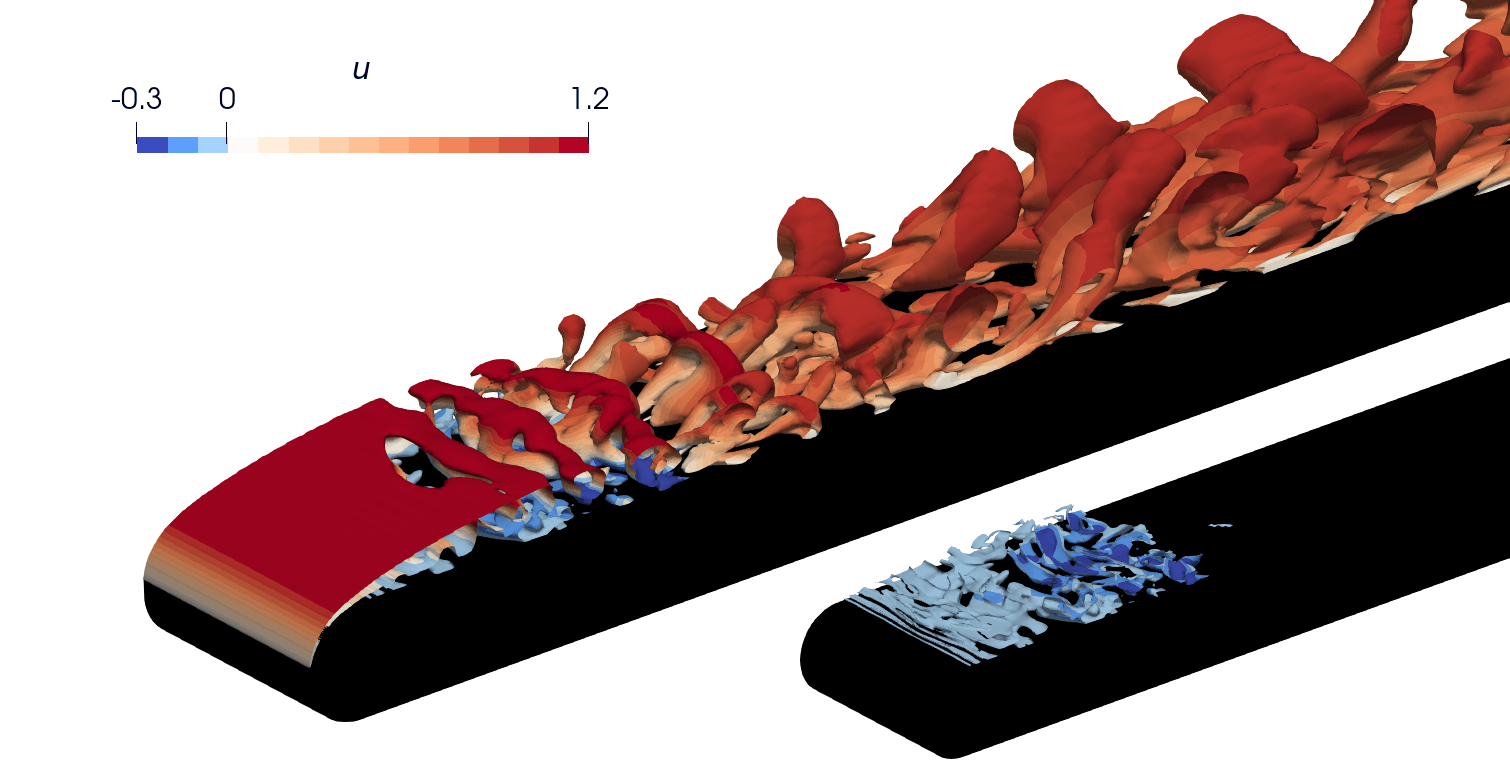}\mylab{-67mm}{17.25mm}{(\ccc)}
	\includegraphics[width=0.45\linewidth, trim={0px 0px 0px 0px},clip]{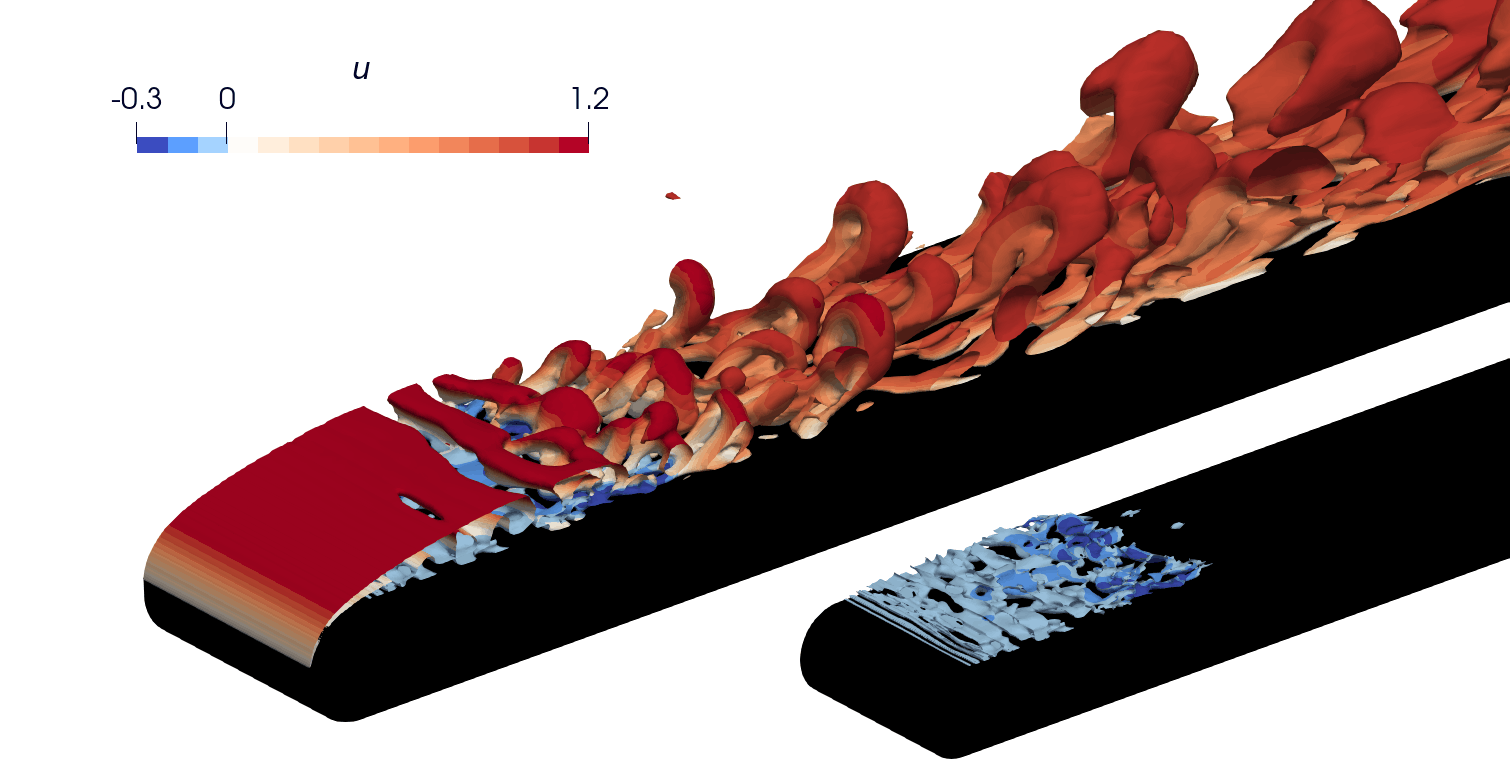}\mylab{0mm}{17.25mm}{(\ddd)}
	\caption{Scale-resolving simulations of the transitional flow over a rounded leading-edge flat plate at medium turbulence intensity $I = 2.3\%$. Iso-surfaces of the Q-criterion \citep{Q} $Q = 0.15$, coloured with streamwise velocity for the various dynamic RANS models:  (a) dynamic $k-\varepsilon$ model on T-Flows, (b) dynamic $k-\varepsilon$ model on OpenFOAM, (c) dynamic $k-\omega$ model on OpenFOAM, (d) zero-decay dynamic $k-\omega$ model on OpenFOAM. Inset plots isolate the reverse boundary layer within the recirculating bubble.}
	\label{fig:Mexican_IFR}
\end{figure}

\begin{figure}
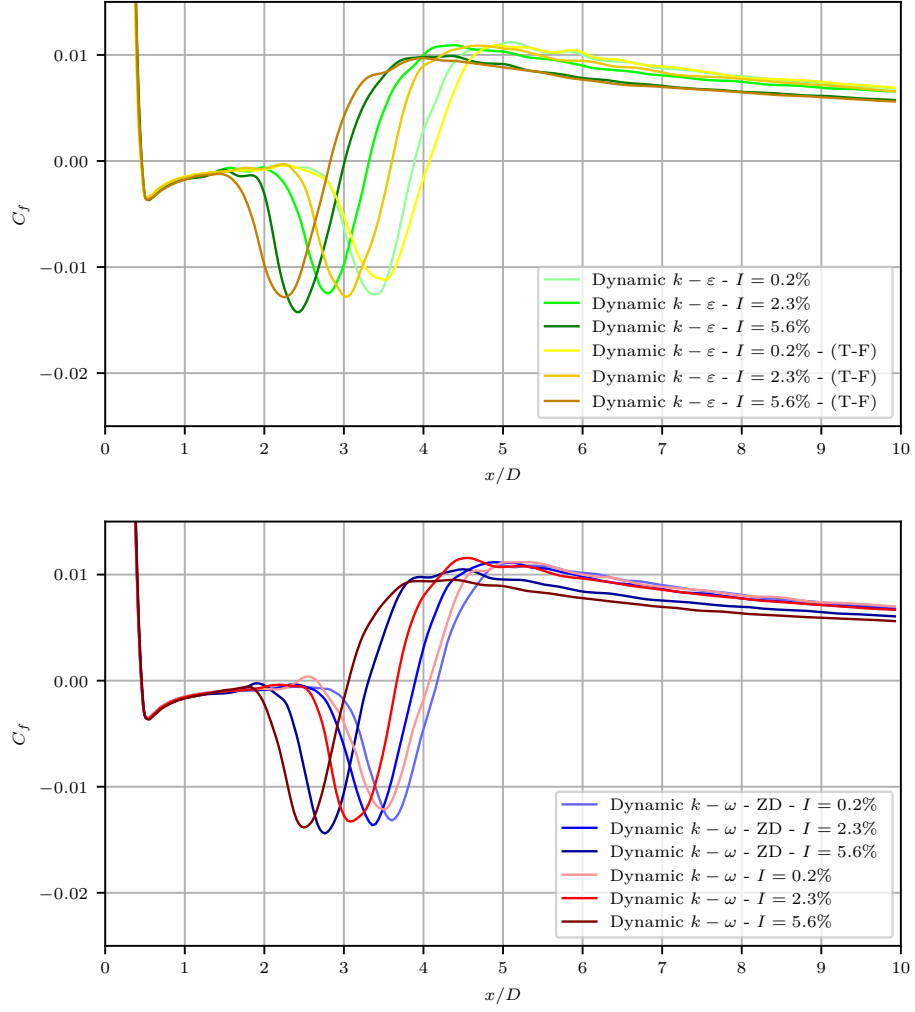

	\centering
	%\makebox[\textwidth][c]{\scalebox{0.9}{\input{Images/Cf_DynamicModelComparison.pgf}}}
	\makebox[\textwidth][c]{\scalebox{0.9}{\input{Images/kEps_T3L_Cf.pgf}}}
	\makebox[\textwidth][c]{\scalebox{0.9}{\input{Images/kOmega_T3L_cf.pgf}}}	
	\caption{Average friction coefficient for the scale-resolving simulations of the transitional flow over a rounded leading-edge flat plate, performed with the various dynamic RANS implementations by varying the free-stream turbulence intensity from low to high. The influence of the Navier-Stokes solver, OpenFoam and T-Flows (T-F), is shown in the top panel by using the dynamic $k-\varepsilon$ model. The effect of the zero-decay approach (ZD) is shown in the bottom panel by using the dynamic $k-\omega$ model. Finally, the influence of the baseline closure, $k-\varepsilon$ vs $k-\omega$, can be appreciated by comparing the top and bottom panels.}
	\label{fig:Mexican_cf}
\end{figure}

Figure \ref{fig:Mexican_IFR} exhibits some instantaneous flow patterns, which immediately demonstrate a very similar qualitative behaviour for all dynamic RANS models, as all the flow features of the unsteady 3D solution are replicated almost identically. In particular, the separation of the laminar boundary layer towards the end of the rounded leading edge and the development of Kelvin-Helmholtz vortices spanning the whole width of the domain are similarly captured by all the approaches. Also the stretching and tilting of these spanwise vortices that eventually leads to hairpin-like structures and velocity streaks that dominate the downstream boundary layer is a common flow feature reproduced by all the approaches. Overall, it appears that the dynamic procedure flattens the differences between the baseline closures and solvers. In particular, the $k-\varepsilon$ formulation translates seamlessly from T-Flows (panel (a) of figure \ref{fig:Mexican_IFR}) to OpenFOAM (panel (b) of figure \ref{fig:Mexican_IFR}), but it also seems that the solution remains equivalent by changing the underlying closure to the $k-\omega$ model (panel (c) of figure \ref{fig:Mexican_IFR}). Furthermore, the similarities then persist when applying the zero decay modification proposed in \S\ref{subsec_zerodecay} (panel (d) of figure \ref{fig:Mexican_IFR}).

A more quantitative view of the solution is given by the evolution of the average friction coefficient $c_f$ along the flat plate. In figure \ref{fig:Mexican_cf} the average $c_f$ profiles from simulations using the dynamic closures under different conditions of free stream turbulence are shown. As anticipated by the instantaneous realizations, all dynamic RANS models capture the laminar separation where $c_f$ becomes negative, transition where $c_f$ reaches its peak negative values, and reattachment of the flow where $c_f$ becomes positive. Also the dependence of these phenomena on the free stream turbulence conditions consisting in a upstream shift of all the phenomena with the free-stream turbulence intensity is captured by all the approaches. Although there is a small spread in the streamwise positions of these physical features of the flow, the profiles remain remarkably similar in their behaviour, and corroborate the idea that the dynamic procedure smooths out the differences between the baseline closures and solvers. For this reason, only the dynamic $k-\omega$ model with the zero-decay approach will be addressed in the following.

\subsection{Dynamic model coefficient}\label{subsubsec_t3lcmu}

\begin{figure}
	\centering
	\includegraphics[width=1.\linewidth, trim={0px 0px 0px 0px},clip]{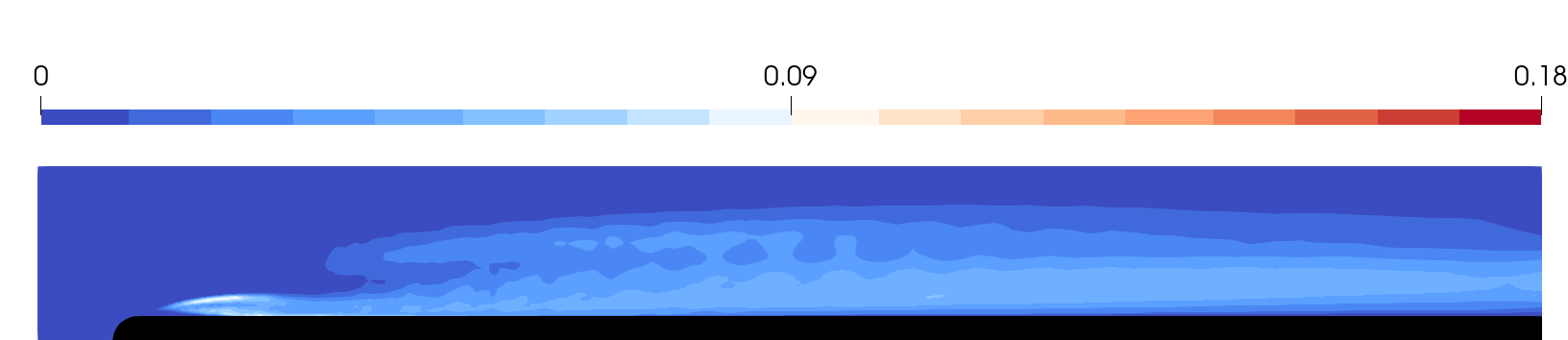}
	\caption{Average field of the model coefficient $\langle c_\mu \rangle$, for the simulation of the transitional flow over a rounded leading-edge flat plate performed with the zero-decay dynamic $k-\omega$ model at medium turbulence intensity $I = 2.3\%$.}
	\label{fig:cmu}
\end{figure}

The mean behaviour of the model coefficient $c_\mu$ is a valuable indicator of the ability of the dynamic procedure to adapt to the local flow characteristics and to the local numerical resolution adopted. Figure \ref{fig:cmu} shows the average distribution of the coefficient for the simulation at medium free stream turbulence $I = 2.3\%$ using the zero-decay dynamic $k-\omega$ closure. It is clear that the procedure keeps $c_\mu$ at very low values along the leading edge, which allows the laminar behaviour of the early shear layer to establish. However, the field then experiences a progressive but swift increase that reflects the onset of transition, which is consequently followed by the reattachment of the flow. The behaviour of the coefficient is also coherent with the reattached boundary layer that continues downstream (and also the reverse boundary layer within the recirculation bubble). The fact that momentum transport is dominated by viscous diffusion in the near-wall region is taken into account by decreasing values of $c_\mu$ approaching the wall, whereas the higher magnitude of $c_\mu$ in the bulk of the expanding boundary layer reflects unresolved turbulent mixing becoming relevant. Finally, the model coefficient decreases again to negligible values away from the boundary layer where the resolved field is laminar and free-stream turbulence is uniquely carried by the transported quantities $k_\tau$ and $\omega_\tau$. In conclusion, the dynamic RANS closure successfully pinpoints the transition of the separated flow, and it autonomously adapts to the characteristics of very different flow regions, by having the turbulent viscosity active in fully turbulent regions while restraining it in laminar areas. The suppression of the eddy viscosity in laminar regions is not compromised by the high shear rates characteristic of the early shear layer or the very near-wall region, and it does not require the use of any damping functions.

\begin{figure}
	\centering
	\includegraphics[width=1.\linewidth, trim={0px 0px 0px 0px},clip]{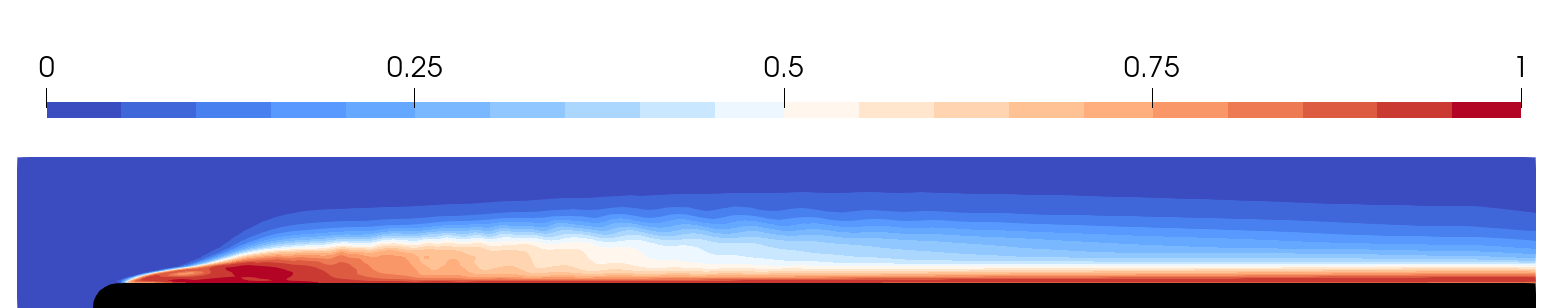}
	\caption{Ratio between the resolved and total kinetic energy, $\mathcal{SR}$, for the simulation of the transitional flow over a rounded leading-edge flat plate performed with the zero-decay dynamic $k-\omega$ model at medium turbulence intensity $I = 2.3\%$.}
	\label{fig:sr_ratio}
\end{figure}

By adapting to the physics of the different flow regions, the behaviour of the model coefficient also enables scale-resolving features. Indeed, the dynamic procedure (\ref{eq_Lilly}) locally adapts the value of the model coefficient $c_\mu$ in order to reproduce the physics of the different flow regions, by capturing flow structures that are supported by the computational grid and time step. To evaluate the scale-resolving degree allowed by the spatial and temporal resolution adopted and enabled by the dynamic procedure, we consider the ratio between resolved to the total kinetic energy,
\begin{equation}
\mathcal{SR} = \frac{\langle \langle u_i \rangle_\tau' \langle u_i \rangle_\tau' \rangle/ 2}{k_\tau + \langle \langle u_i \rangle_\tau' \langle u_i \rangle_\tau' \rangle / 2}
\end{equation}
As shown in figure \ref{fig:sr_ratio}, this ratio is vanishingly small in the free-stream regions where the resolved solution is smooth and free-stream turbulence is exclusively carried by the transported quantities $k_\tau$ and $\omega_\tau$. On the other hand, this ratio increases by moving in the turbulent bulk of the flow reaching values close to unity within the turbulent core of the separation bubble, in the turbulent shedding region and in the fully-developed boundary layer. Hence, a scale-resolving simulation is found to be supported by the spatial and temporal resolution adopted and enabled by the dynamically-adapted model coefficient $c_\mu$.

\subsection{Dynamic RANS performances with respect to state-of-art closures}\label{subsubsec_t3lkiles}

The dynamic RANS performances expressed by the zero-decay dynamic $k-\omega$ model are analysed here by comparing its results with those of state-of-art closures. In particular, the standard $k-\omega$ model will be considered to address the improvements provided by the dynamic procedure to its baseline closure. On the other hand, the hybrid $k-\omega$ SST-DES model \cite{SSTDES} will be considered to address the performance of the dynamic RANS approach in reproducing the scale-resolving features of the solution.

\begin{figure}
	\centering
	\includegraphics[width=0.45\linewidth, trim={0px 0px 0px 0px},clip]{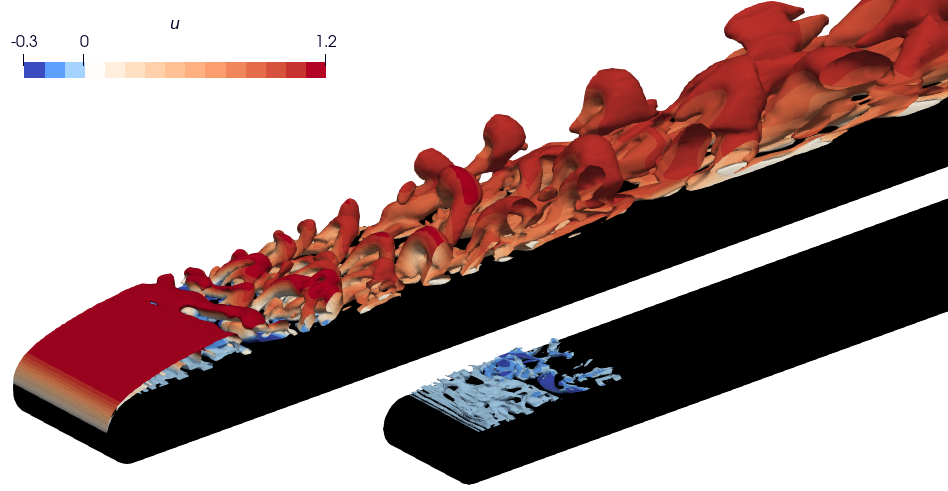}\mylab{-67mm}{17.25mm}{(\aaa)}
	\includegraphics[width=0.45\linewidth, trim={0px 0px 0px 0px},clip]{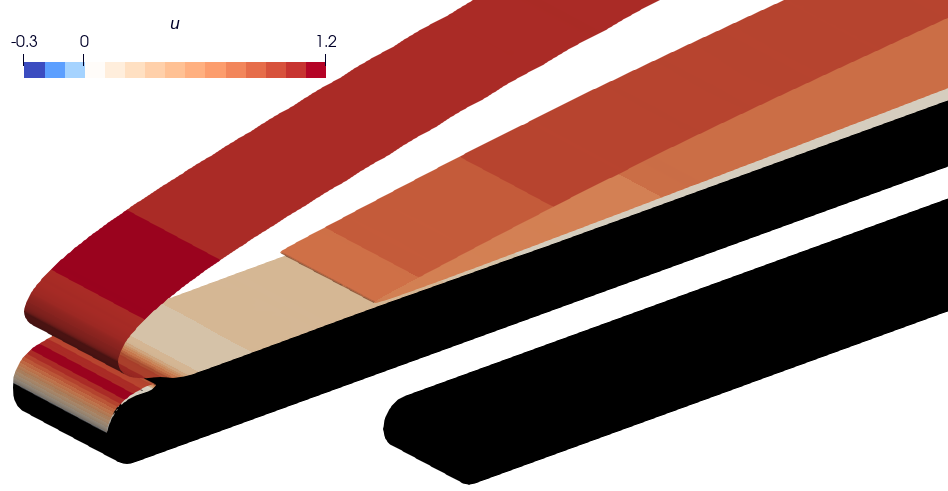}\mylab{0mm}{17.25mm}{(\bbb)}
	\includegraphics[width=0.45\linewidth, trim={0px 0px 0px 0px},clip]{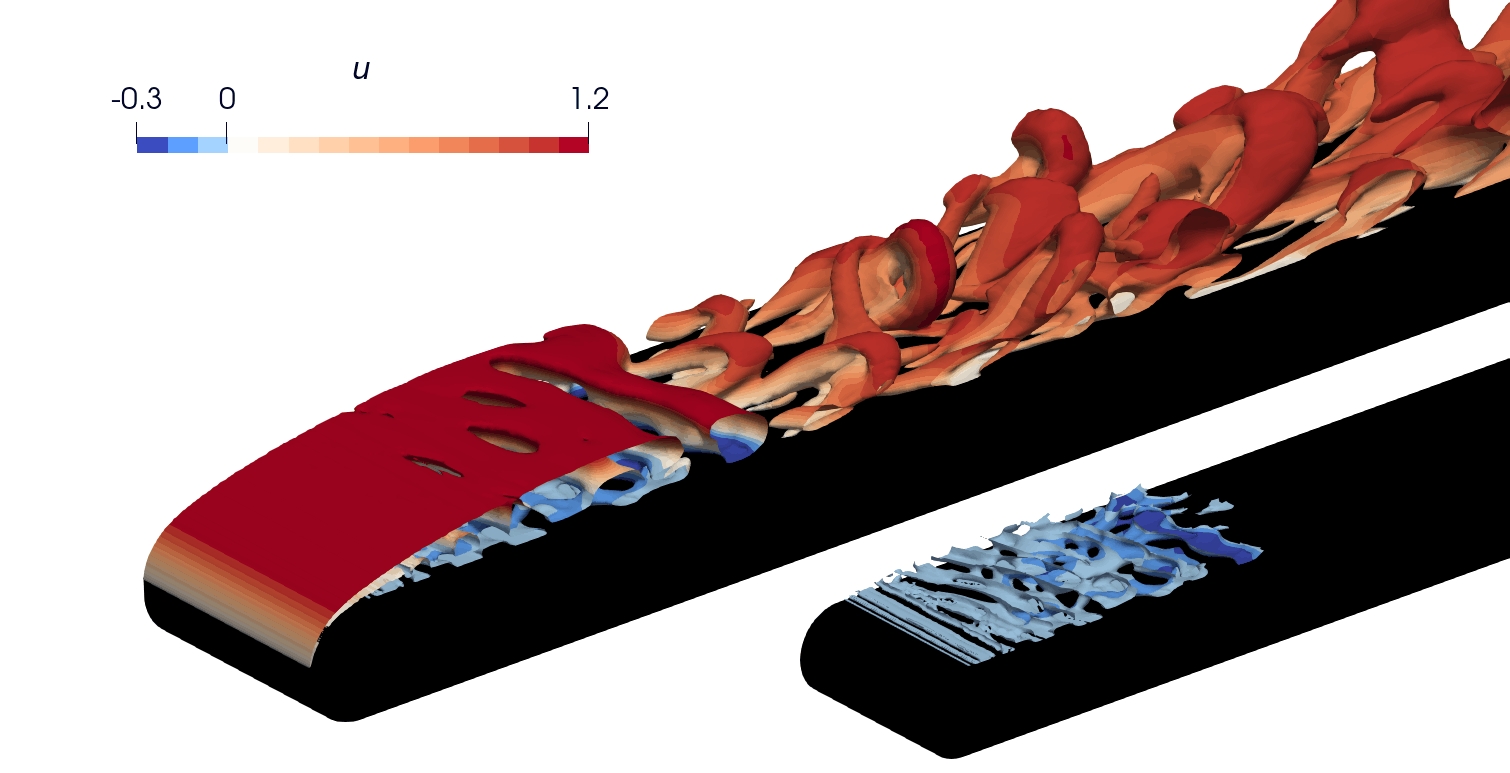}\mylab{-67mm}{17.25mm}{(\ccc)}
	\caption{Simulations of the transitional flow over a rounded leading-edge flat plate at medium turbulence intensity $I = 2.3\%$. Iso-surfaces of the Q-criterion, $Q = 0.15$, coloured with streamwise velocity for the various approaches based on the $k-\omega$ model:  (a) Zero-decay dynamic $k-\omega$ model, (b) standard $k-\omega$ model, (c) $k-\omega$ SST-DES model. Inset plots isolate the reverse boundary layer within the recirculating bubble.}
	\label{fig:kos_IRF}
\end{figure}

The qualitative differences between the modelling approaches are shown in figure \ref{fig:kos_IRF} where the instantaneous flow realizations are depicted. As anticipated in \S\ref{subsubsec_t3ldyndyndyn}, the dynamic approach shown in figure \ref{fig:kos_IRF}(a) successfully replicates the main features of the fully unsteady 3D solution of the transitional flow about a rounded leading edge flat plate. As shown in figure \ref{fig:kos_IRF}(c), the $k-\omega$ SST-DES model is also capable of reproducing the unsteadiness of the flow, although the solution is visibly different from the former. In particular, the transition in the shear layer reproduced by the hybrid model develops more slowly, with a subsequent delay in the phenomena that follow. Nevertheless, all the main features are eventually recovered. On the other hand, the solution yielded by the standard $k-\omega$ formulation is completely stationary and two-dimensional as shown figure \ref{fig:kos_IRF}(b). The separated shear layer is included in the solution, but all the other features of the flow --which the dynamic and hybrid approaches prove can be generously supported by the spatial resolution-- are completely absent. These differences clearly highlight how the use of hybrid approaches can have a significant impact on the underlying statistical closure's solution and, remarkably, that the same improvement can be achieved by the sole adoption of the dynamic procedure.

\begin{figure}
	\centering
	\makebox[\textwidth][c]{\scalebox{0.9}{\input{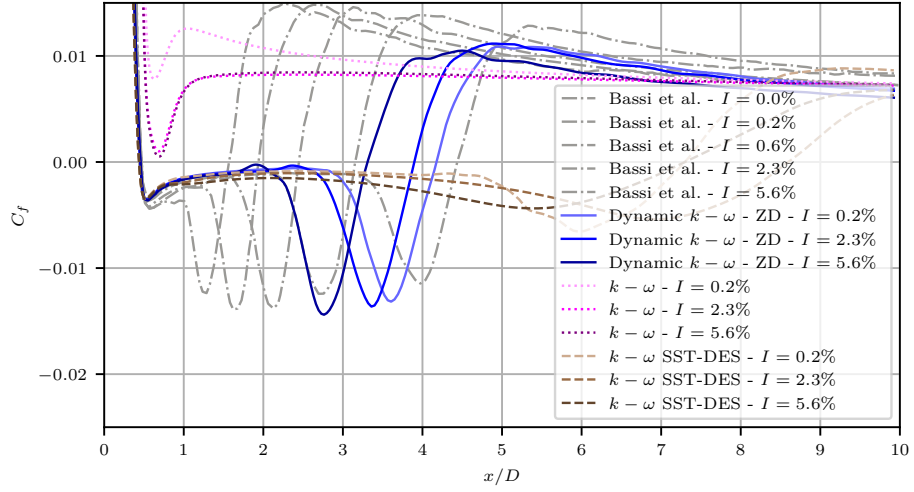}}}
	\caption{Average friction coefficient for the simulations of the transitional flow over a rounded leading-edge flat plate performed with the various approaches based on the $k-\omega$ model, compared to the high-fidelity implicit-LES simulations performed by \citet{Bassi2020}.}
	\label{fig:kos_cf}
\end{figure}

In order to quantitatively assess the differences between the modelling approaches in reproducing the laminar separation, transition, and reattachment of the T3L flow, figure \ref{fig:kos_cf} reports the average profiles of the friction coefficient $c_f$ retrieved from the present simulations. The dynamic RANS approach captures not only the aforementioned features, but the dependence of the solution on the modelled free-stream turbulence as well, with the separation bubble shrinking as the turbulence intensity $I$ increases. The underlying $k-\omega$ closure, on the other hand, fails to reproduce any of the flow's characteristic phenomena. Due to its well-known tendency to overpredict eddy viscosity in laminar and transitional regions, it predicts a fully turbulent boundary layer too early along the solid body, which consequently does not separate. Moreover, the unmodified $k-\omega$ model also responds incorrectly to the variation of the modelled free-stream turbulence intensity --- this was not the case for the standard $k-\varepsilon$ model in our previous study \cite{Noi}, which suggests the issue could be related to the $k-\omega$ formulation's sensitivity to the free stream values of $\omega$. In any case, it is clear that the physics of transitional separating and reattaching flows poses significant difficulties for the unmodified $k-\omega$ model. However the adoption of the dynamic procedure allows it to overcome these limitations, rendering it a suitable solution for such cases.

The evolution of $c_f$ extracted from the simulations with the hybrid $k-\omega$ SST-DES model confirms that it is also capable of capturing the characteristic features of the solution at hand, but a comparison with the reference ILES results \citep{Bassi2020} indicates that the reproduction is much less faithful than with the dynamic RANS approach. Indeed, the latter reproduces the shape of the profiles, the intensity of their peaks and the reattachment length much better than the DES solution. Furthermore, the hybrid $k-\omega$ SST-DES model is found to be responsive to the variation of the free-stream turbulence intensity $I$ imposed through the inlet values of $k$ and $\omega$ but in a less physical way with respect to the dynamic RANS approach. Indeed, the physically coherent upstream shift of all the phenomena with the free-stream turbulence intensity $I$ captured by the dynamic RANS closure is missed by the DES approach which predicts a non-monotonic dependence of the location of reattachment with $I$.

\subsection{Robustness to a coarsening of the temporal resolution}\label{subsubsec_t3lCFL}

Hybrid approaches extending statistical models with scale-resolving closures or capabilities can offer richer (and, usually, more accurate) reproductions of flow solutions than the underlying RANS formulations alone, often on the very same grids required for the RANS simulations. However, they commonly require very fine temporal resolutions to function properly (classically $CFL < 1.5$), which places them back at a disadvantage relative to the purely statistical approaches in terms of computational cost. This is one of the main reasons why their practical adoption is limited, making improved robustness to temporal resolution one of the most critical assets for hybrid approaches to pursue. To that end, a temporal resolution sensitivity study is conducted here to benchmark the dynamic $k-\omega$ closure against the hybrid $k-\omega$ SST-DES formulation. The medium free stream turbulence intensity $I = 2.3\%$ is considered and the original time step, tuned to maintain a condition $CFL<1.5$, is coarsened by a factor 3, 10, 50, and 100.

\begin{figure}
	\centering
	\includegraphics[width=0.45\linewidth, trim={0px 0px 0px 0px},clip]{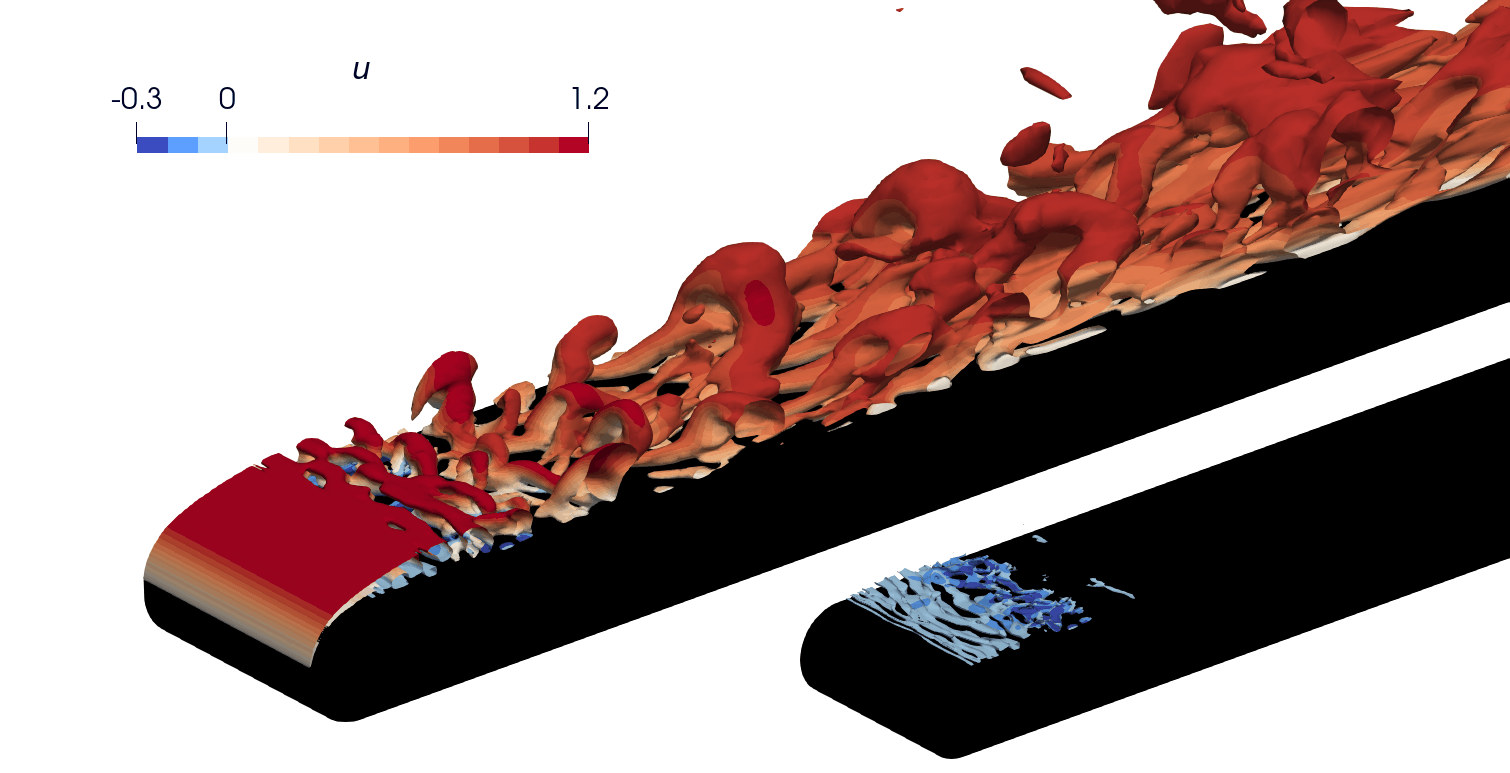}\mylab{-65mm}{17.25mm}{(\aaa)}
	\includegraphics[width=0.45\linewidth, trim={0px 0px 0px 0px},clip]{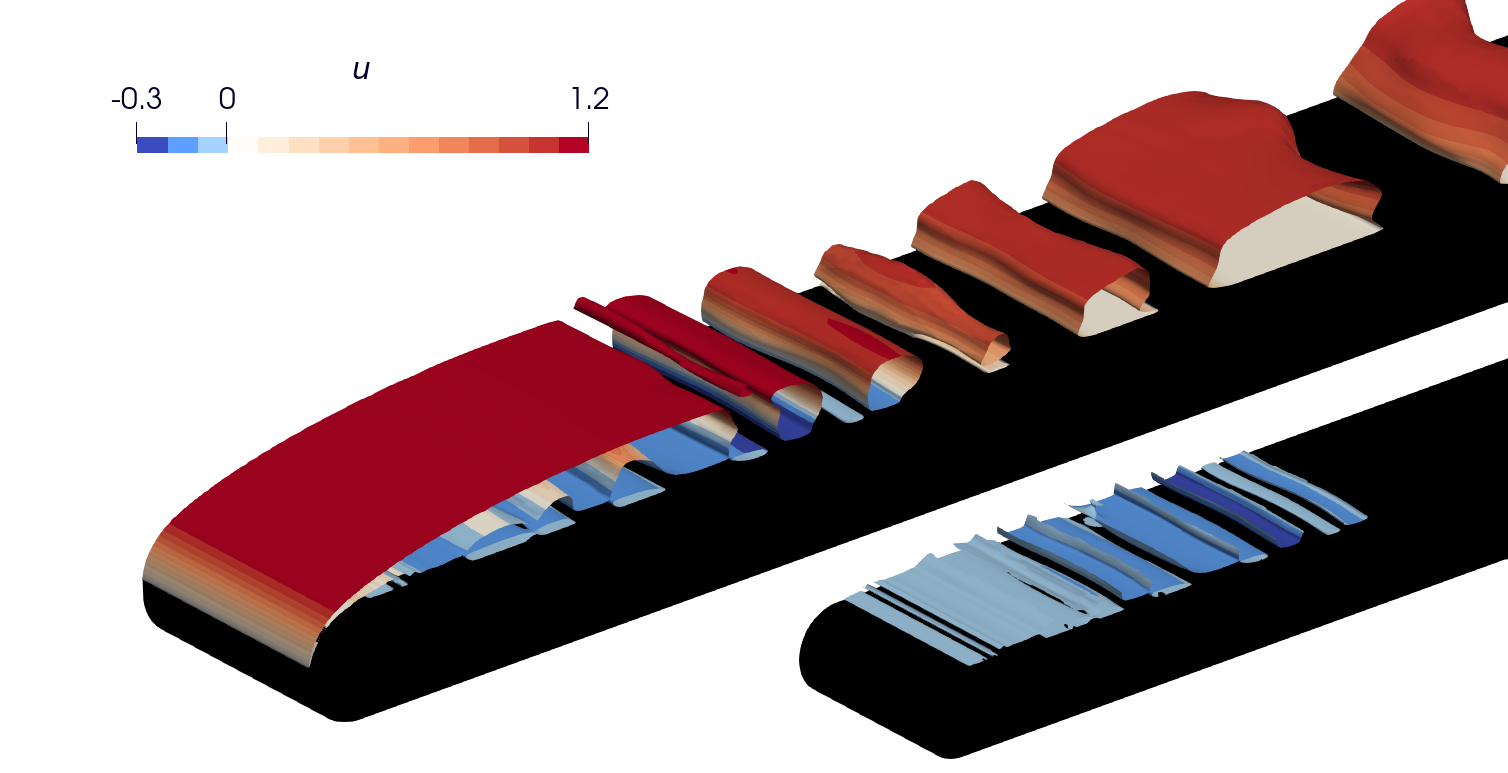}\mylab{1mm}{17.25mm}{(\aaa$'$)}
	
	\includegraphics[width=0.45\linewidth, trim={0px 0px 0px 0px},clip]{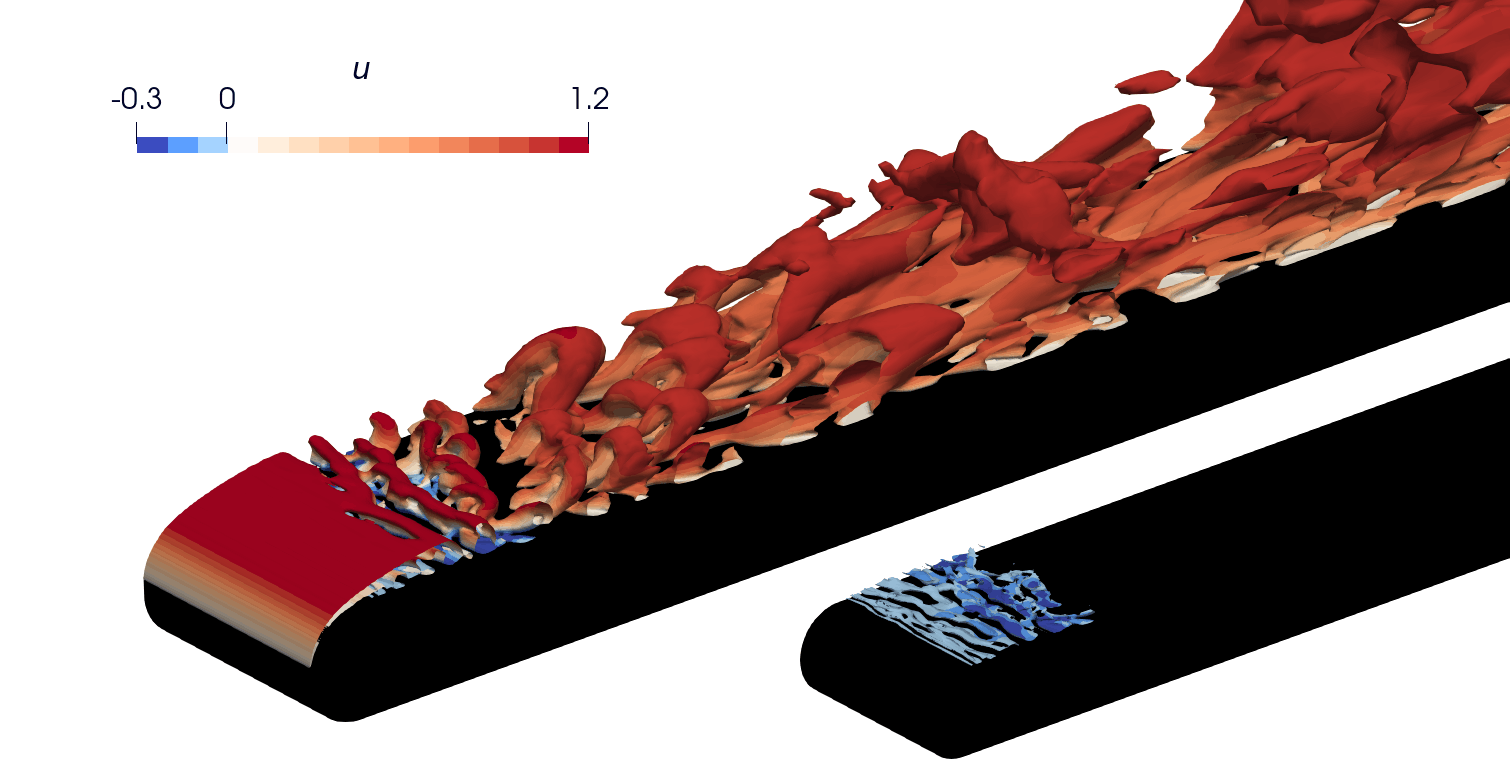}\mylab{-65mm}{17.25mm}{(\bbb)}
	\includegraphics[width=0.45\linewidth, trim={0px 0px 0px 0px},clip]{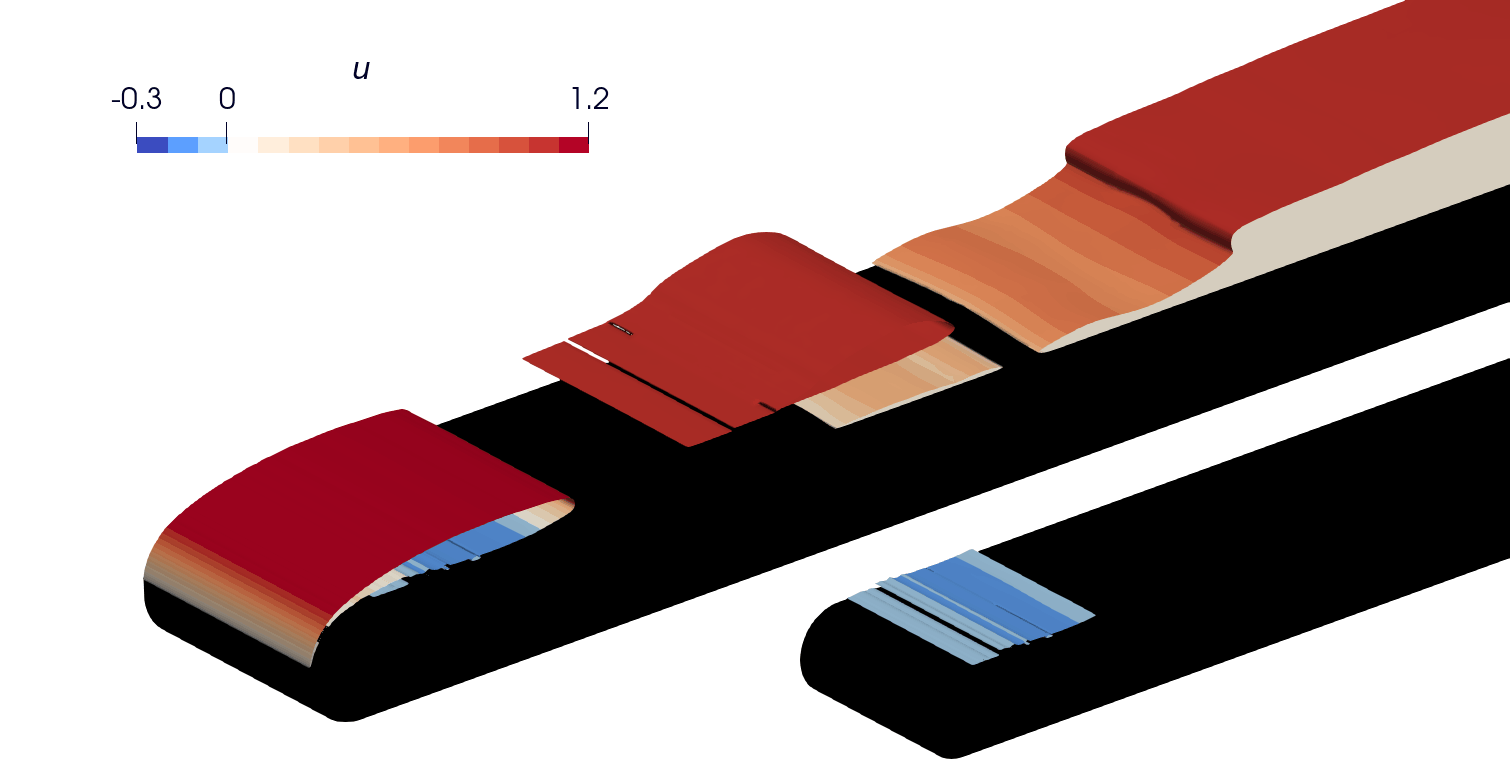}\mylab{1mm}{17.25mm}{(\bbb$'$)}
	
	\includegraphics[width=0.45\linewidth, trim={0px 0px 0px 0px},clip]{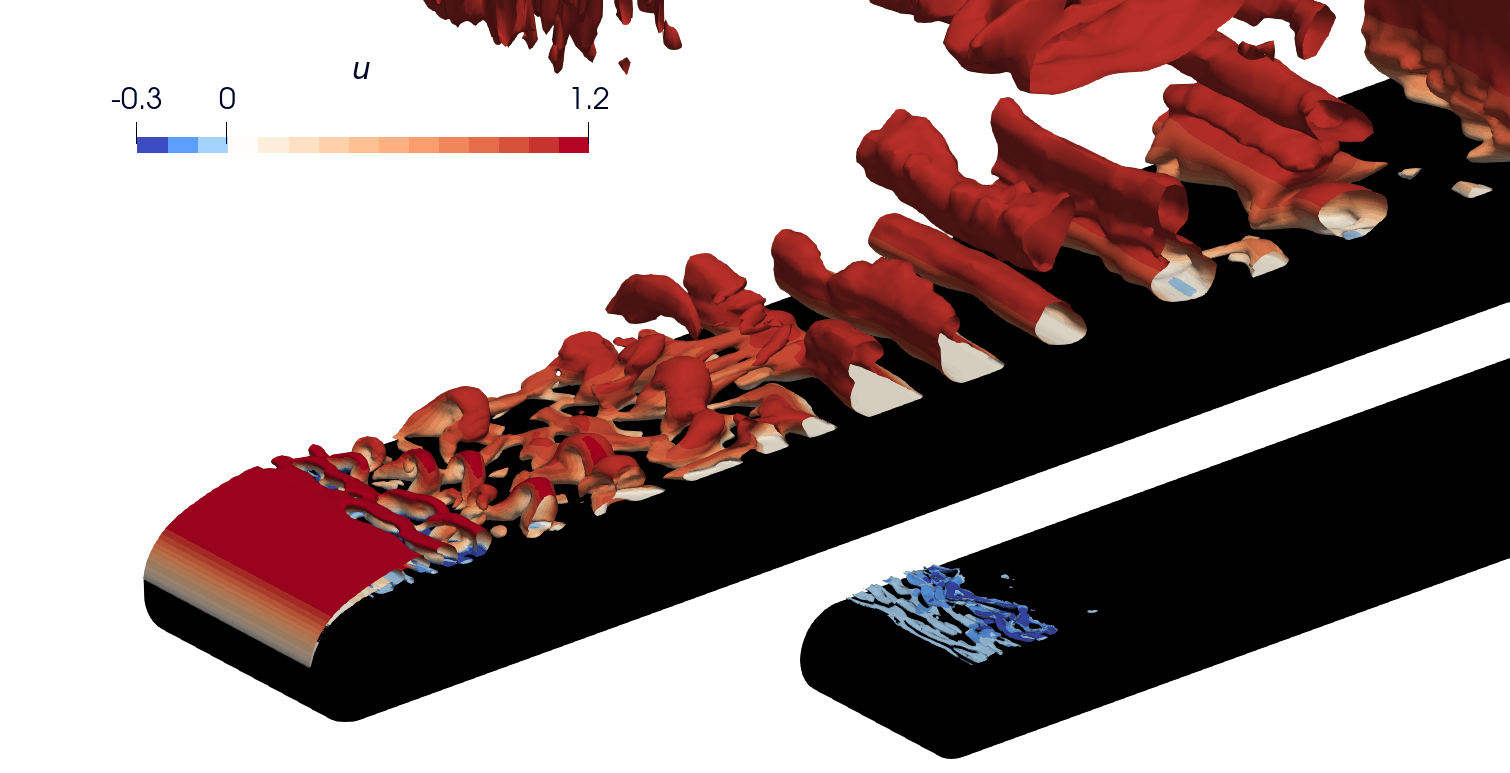}\mylab{-65mm}{17.25mm}{(\ccc)}
	\includegraphics[width=0.45\linewidth, trim={0px 0px 0px 0px},clip]{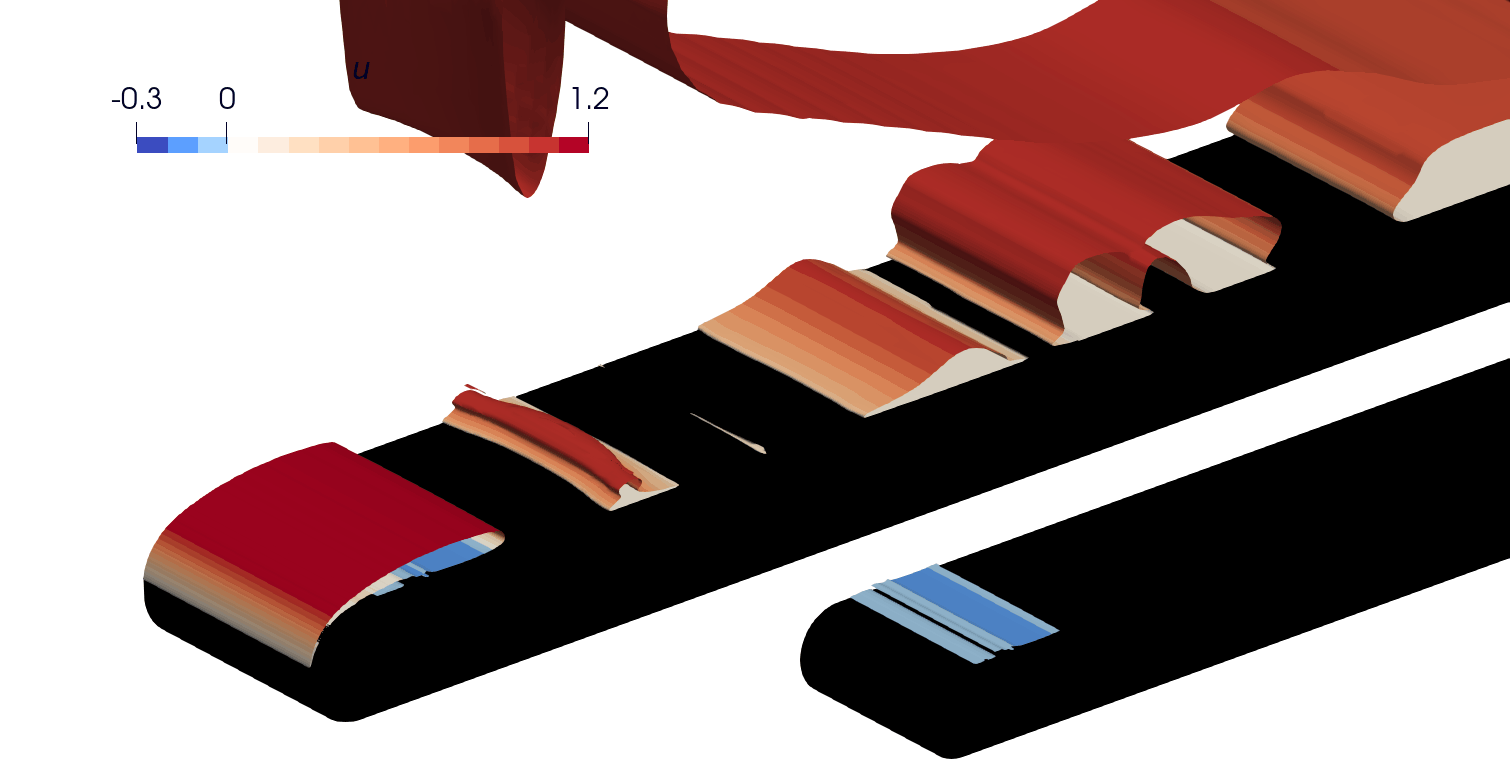}\mylab{1mm}{17.25mm}{(\ccc$'$)}
	\caption{Simulations of the transitional flow over a rounded leading-edge flat plate. Iso-surfaces of the Q-criterion, $Q = 0.15$, colored with streamwise velocity for the various  simulations performed at varying temporal resolutions at medium turbulence intensity $I = 2.3\%$, with the zero-decay dynamic $k-\omega$ model (left panels) and the hybrid $k-\omega$ -SST-DES model (right panels). Inset plots isolate the reverse boundary layer within the recirculating bubble. From top to bottom, the time step is coarsened by a factor 10, 50 and 100.}
	\label{fig:CFL_IFR}
\end{figure}

The evolution of the behaviour of the instantaneous flow realisations, shown in figure \ref{fig:CFL_IFR}, is particularly revealing. While both models continued to yield qualitatively comparable results increasing the time step by a factor of 3 (not shown), the solution starts to degrade heavily for the DES approach starting at 10 times the reference time step, as shown in figure \ref{fig:CFL_IFR}(a'). A fully two-dimensional solution is then reproduced at 50 times the reference time step, as shown in figure \ref{fig:CFL_IFR}(b'). On the contrary, the solution for the dynamic $k-\omega$ model does not start to show evident signs of deterioration of the solution until the time step reaches 100 times the reference value, although a three-dimensional flow pattern is still visible, as shown in figure \ref{fig:CFL_IFR}(a-c).

\begin{figure}
	\centering
	\makebox[\textwidth][c]{\scalebox{0.9}{\input{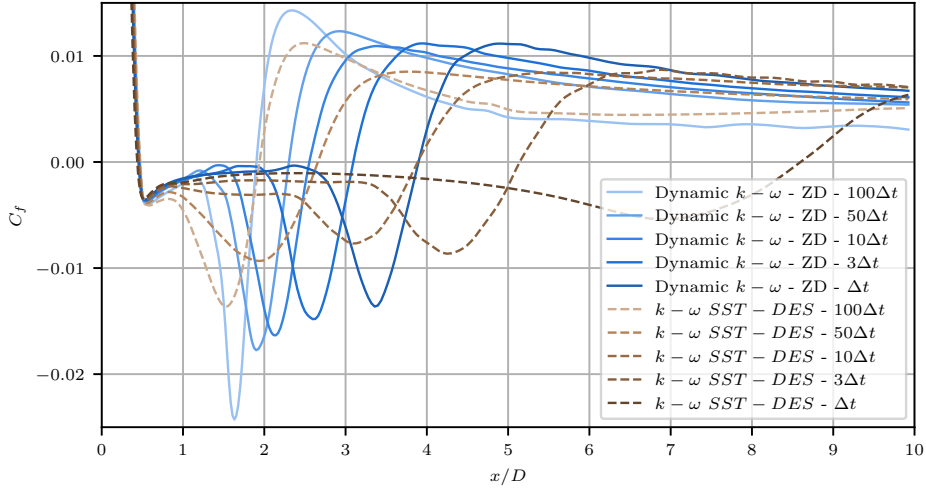}}}
	\caption{Average friction coefficient for the simulations of the transitional flow over a rounded leading-edge flat plate. Simulations performed at medium turbulence intensity $I = 2.3\%$, at varying temporal resolutions, with the zero-decay dynamic $k-\omega$ model and the $k-\omega$ SST-DES model.}
	\label{fig:cfL}
\end{figure}

The response of the average friction coefficient profiles to the changing temporal resolution is reported in figure \ref{fig:cfL}. The recirculating bubble shrinks for the dynamic RANS simulations as the time step increases, because fewer turbulent fluctuations can manifest through temporal oscillations of the solution and the eddy viscosity increases as a consequence, resulting in an earlier reattachment of the flow. A comparable mechanism affects the DES simulations where a coarser temporal resolution leads to a solution that is determined to a greater extent by the RANS mode of the closure, which results again in a higher eddy viscosity. The upstream shift of reattachment comes hand in hand with a deeper trough of the friction coefficient, provided that it occurs early enough for the streamwise velocity fluctuations induced by the Kelvin-Helmhotz vortices to make an appreciable difference --- the effect is most evident at the coarsest resolution levels. However, the most important sign of the solution collapsing is that the intensity of the friction coefficient in the reattached boundary layer is compromised for both approaches starting at 100 times the reference time step.

Overall, the dynamic $k-\omega$ model appears to be much more robust than the hybrid $k-\omega$ SST-DES to the coarsening of temporal resolution, and it remains reliable at remarkably coarse discretisation levels up to $CFL \approx 75$. Moreover, the dynamic procedure allows the solution to hold on to a wide range of turbulent scales throughout the temporal resolutions considered, whereas the DES model starts losing its scale-resolving capabilities at a temporal resolution over an order of magnitude finer.

%%%%%%%%%%%%%%%%%%%%%%%%%%%%%%%%%%%%%%%%%%%%%%%%%%%%%%%%%%%%%%%%%%%%%%%%%%%%%%%%%%%%%

\section{Flow over a NACA 4412 wing section at near-stall conditions}\label{subsec_nacaresults}

The flow over a NACA 4412 wing section at near-stall conditions is used in this section to investigate the model's ability to predict the onset of separation under strong adverse pressure gradients at high Reynolds numbers, when coupled with wall functions which are necessary when using very-coarse wall-normal resolutions, characteristic of industrial applications. 

The simulations employ the zero-decay dynamic $k-\omega$ closure. Numerical predictions are initially validated against the experimental data of \citet{Wadcock1979}, which provides the baseline for mean velocity profiles and surface pressure distributions at near-stall conditions. To supplement the experimental validation, the model is further benchmarked on the same computational grid against the canonical RANS closure $k$--$\omega$ SST. This comparative strategy facilitates a cross-verification between experimental evidence and established industrial standards, thereby isolating the predictive capabilities of the proposed dynamic procedure in high Reynolds-number separated flows. Consequently, the comparison focuses not merely on a direct quantitative match of mean coefficients, but also on the qualitative and structural response of the dynamic model in high-Reynolds-number regimes solved by using very-coarse wall-normal spatial resolution.

Flow statistics are computed by averaging both in time and along the homogeneous spanwise direction. Specifically, 30 temporal samples are collected with a time interval $c/U_0$ where $c$ is the chord of the wing, after a statistically stationary state is reached.

\subsection{Instantaneous flow pattern}

\begin{figure}
	\centering
	\includegraphics[width=1.\linewidth, trim={0px 0px 0px 0px},clip]{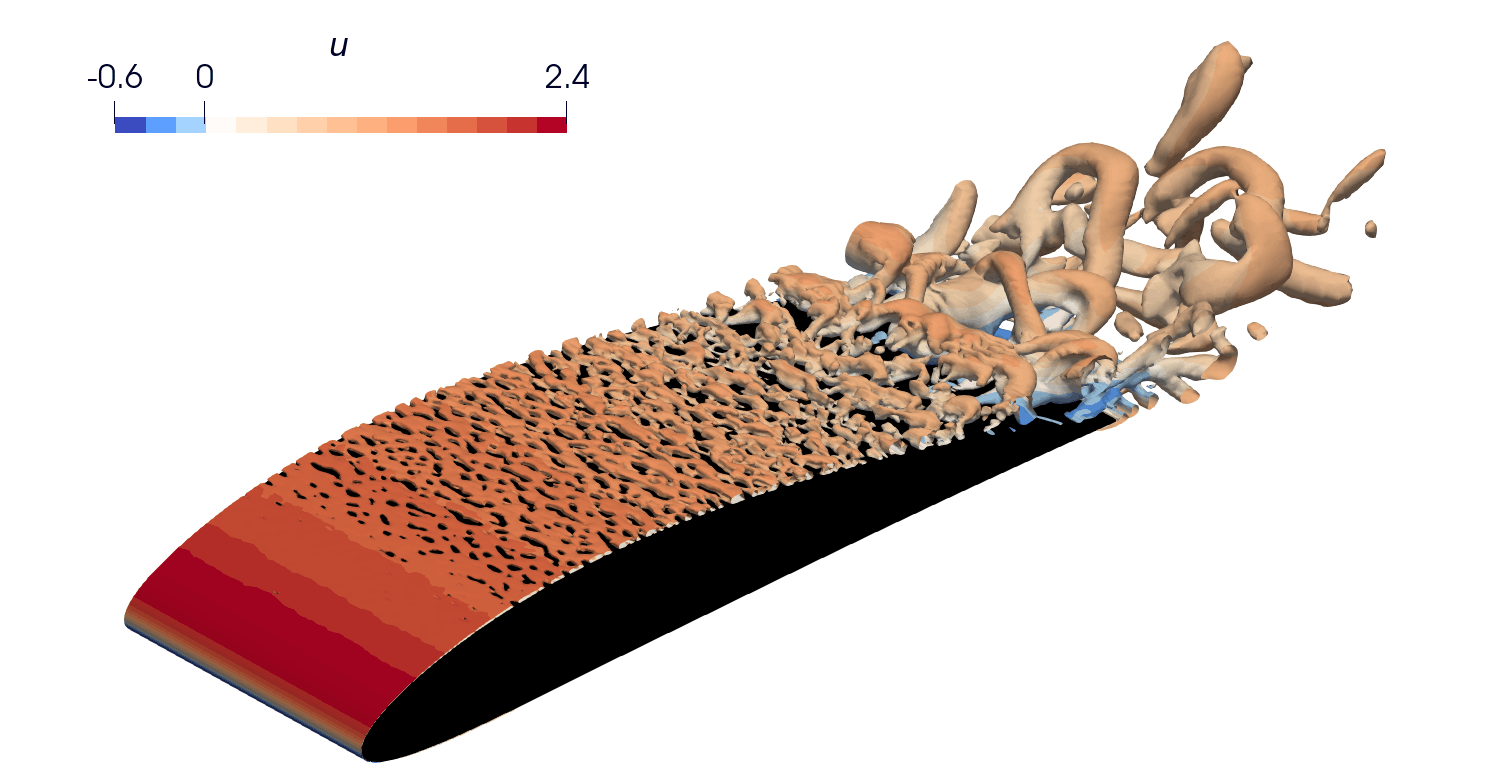}
	\caption{Scale-resolving simulation of the flow over a NACA 4412 at $Re = 1.52 \cdot 10^6$ and angle of attack $\alpha = 13.87^\circ$. Iso-surfaces of the Q-criterion $Q =20000$, coloured with streamwise velocity for the zero-decay dynamic $k-\omega-ZD$ model.} 
	\label{fig:airfoil_inst}
\end{figure}

The global flow topology and the turbulent structures captured by the dynamic $k-\omega$ model are presented in figure \ref{fig:airfoil_inst} where an instantaneous flow realization is reported. The boundary layer remains predominantly attached over the suction surface, with the onset of turbulent separation occurring at $x/c \approx 0.74$ that is close to the experimental value $x/c \approx 0.85$ of \citet{Wadcock1979}. A recirculating region is developed from which turbulent structures are detached and advected downstream. Despite the inherent resolution constraints related to the wall-modeled approach employed (the average thickness of the first layer of cells in the suction side is 50 friction units), the dynamic RANS model successfully captures relevant scale-properties of the solution such as the flow recirculation unsteadiness and the coherent structures populating the attached boundary layer and the wake. Conversely, the $k-\omega$ SST formulation yields a strictly steady-state solution, inherently suppressing the transient dynamics resolved by the dynamic approach (not shown for brevity).

\subsection{Aerodynamic forces}

The predictive capability of the numerical framework is demonstrated here by analyzing the integral aerodynamic forces. As shown in table \ref{tab:forces_comp}, the mean lift coefficient is in good agreement with experimental data \citep{Wadcock1979}. On the other hand, the drag coefficient exhibits a significant underestimation with respect to experimental data \citep{Wadcock1979}. These predictions are however in better agreement with respect to the computation performed with the baseline eddy viscosity framework on the same coarse grid, see again table \ref{tab:forces_comp}. The intensity of the aerodynamic force fluctuations is also evaluated by computing their standard deviation even if no reference data are available for comparison. We measure $\sigma_{C_l} = 0.0161$ and $\sigma_{C_d} = 0.0296$. Both the mean and standard deviation of the lift and drag are the integral result of friction and pressure stresses distributed along the wing profile. The latter are analysed in the following. 

\begin{table}
  \centering
  \begin{tabular*}{\linewidth}{@{\extracolsep{\fill}} lccc }
    \toprule
    Turbulence Model & \(C_l\) & \(C_d\) & \(x_s/c\) \\
    \midrule
    \(k - \omega\) SST          & 1.899 & 0.0218 & 0.972 \\
    Dynamic \(k - \omega\) ZD   & 1.75  & 0.030  & 0.74  \\
    \midrule
    Experimental (Ref.)         & 1.69  & 0.072  & 0.85  \\
    \bottomrule
  \end{tabular*}
\caption{Comparison of time-averaged lift coefficient \(C_l\), drag coefficient \(C_d\) and separation point \(x_s\) for the flow over a NACA 4412 at \(Re = 1.52 \cdot 10^6\) and angle of attack \(\alpha = 13.87^\circ\). The current zero-decay dynamic \(k-\omega\) simulation results are compared against the experimental data of \citet{Wadcock1979} and the numerical solution using the \(k-\omega\) SST turbulence closure.}
\label{tab:forces_comp}
\end{table}

\begin{figure}
	\centering
    \makebox[\textwidth][c]{\scalebox{0.9}{\input{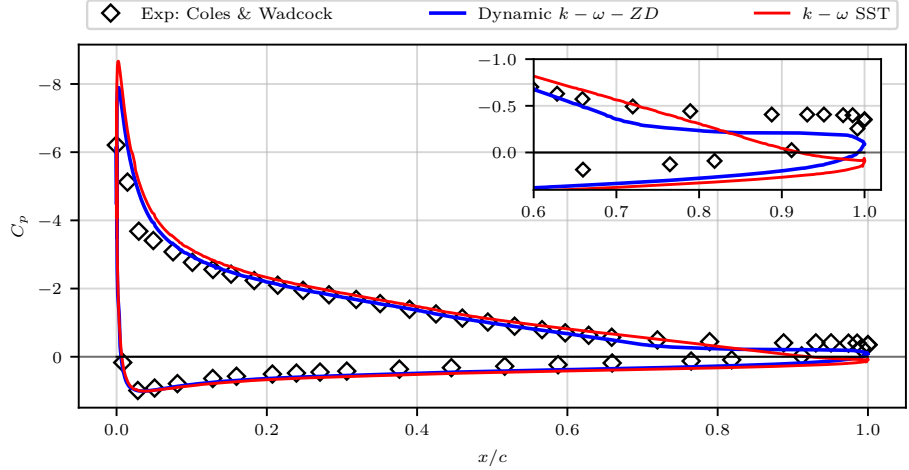}}}
	\caption{Average pressure coefficient for the scale-resolving simulation of the flow over a NACA 4412 wing section at $Re=1.52\cdot 10^6$ and $\alpha=13.87^\circ$. The current zero-decay dynamic $k-\omega$ simulation results are compared against the experimental data of \citet{Wadcock1979} and the numerical solution using the $k-\omega$ SST turbulence closure.}
	\label{fig:Cp_max}
\end{figure}

Figure \ref{fig:Cp_max} reports the distribution of the pressure coefficient $c_p$ obtained from the current dynamic RANS formulation, the experimental investigations and standard computational frameworks. A good agreement of the results from the dynamic RANS model with the experimental results is overall observed. By going more in detail, it is worth addressing the behaviour of the pressure coefficient in the first and last portion of the wing section. Starting from the trailing-edge region, it is known that accurately predicting the pressure plateau over the last portion of the wing ---specifically the last $40\%$ of the chord--- constitutes a primary numerical challenge in near-stall flows \citep{Frere2018}. As shown in figure \ref{fig:Cp_max}, the baseline RANS formulation consistently struggles in this region, predicting an excessively steep pressure recovery that indicates a delay in the onset of flow separation. On the contrary, the proposed dynamic RANS model successfully recovers a more pronounced pressure plateau, demonstrating an alignment with the experimental evidence. Focusing now on the leading edge, figure \ref{fig:Cp_max} highlights that the standard $k-\omega$ SST formulation overpredicts the magnitude of the suction peak. Conversely, the dynamic RANS approach effectively reduces this overestimation, yielding a pressure distribution that is notably closer to the experimental data of \citet{Wadcock1979}.

\subsection{Dynamic model coefficient}

\begin{figure}
	\centering
	\includegraphics[width=0.8\linewidth, trim={0px 0px 0px 0px},clip]{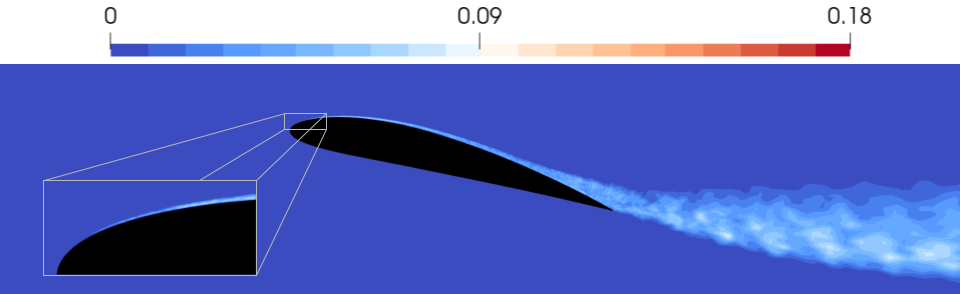}
	\caption{Average behaviour of the dynamic coefficient $c_\mu$, for the simulation of the transitional flow over a NACA4  4412 wing section performed with the zero-decay dynamic $k-\omega$ model. }
	\label{fig:Airfoil_Cmu}
\end{figure}

As documented by \citet{Menter1992}, eddy-viscosity models overpredict unphysically levels of eddy viscosity in the vicinity of the stagnation point.  This occurs even outside the boundary layer, a result that is inconsistent with the expected laminar behavior in such regions.  Such non-physical growth of turbulent quantities is rooted in the Boussinesq hypothesis, which fails in regions dominated by normal strain rates. Specifically, the production of turbulent kinetic energy in regions of vanishing shear strain is proportional to the eddy viscosity and the square of the normal strain rate. Hence, near a stagnation point, the high velocity gradients induce a non-zero production term, thus enforcing a fully turbulent boundary layer immediately downstream of the stagnation point. This behaviour alters the streamwise evolution of the boundary layer from its inception and, by artificially enhancing the momentum transfer towards the wall, confers an increased resilience against the adverse pressure gradient. As quantified in table \ref{tab:forces_comp}, this effect leads the $k-\omega$ SST model to predict an attached boundary layer over almost the entire suction side of the wing.

Conversely, the production of turbulent quantities and the intensity of the eddy viscosity in the dynamic RANS formulation are directly modulated by the dynamic coefficient. Since $c_\mu$ is determined locally based on the resolved scales and the underlying flow physics, it naturally vanishes in regions where turbulent fluctuations are absent or weak such as the stagnation region. As shown in figure \ref{fig:Airfoil_Cmu}, the activation of the dynamic RANS closure is intrinsically linked to the growth of turbulent fluctuations. It occurs in close proximity to the leading edge and aligns remarkably well with the transition envelopes documented in both wall-resolved and wall-modelled LES literature.  Consequently, under a natural-transition regime, the spatial onset of the dynamic coefficient serves as a robust proxy for the transition point identification. This self-regulating behavior represents a significant theoretical advantage over traditional eddy viscosity models. As a matter of facts, the dynamic framework recovers a more sensitive response of the boundary layer to the adverse pressure gradient on the suction side, predicting the onset of turbulent separation as confirmed by experimental data \citep{Wadcock1979}.

\subsection{Velocity profiles}

\begin{figure}
	\centering
    \makebox[\textwidth][c]{\scalebox{0.9}{\input{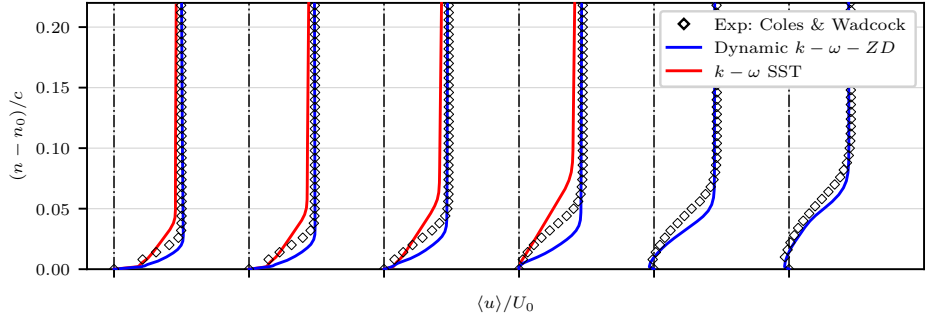}}}
	\caption{Velocity profiles on the suction side of a NACA 4412 airfoil at $Re=1.52\cdot 10^6$ and $\alpha =13.87^\circ$ evaluated at different streamwise positions with respect to the separation point, $(x-x_s)/c = -0.165$, $-0.11$, $-0.055$, $0$, $0.055$ and $0.11$ from left to right. The current zero-decay dynamic $k-\omega$ simulation results are compared against the experimental data of \citet{Wadcock1979} and the numerical solution using $k-\omega$ SST turbulence closure.}
	\label{fig:Velocity_profile_airfoil}
\end{figure}

As a final assessment of the dynamic RANS model performance, we address the mean velocity profiles extracted at different streamwise locations along the suction side. To ensure a consistent comparative framework despite the different locations of the boundary layer separation characterizing the numerical and experimental datasets, the average velocity profiles are compared at the same streamwise distance from the separation point. As shown in figure \ref{fig:Velocity_profile_airfoil}, a reasonable agreement between the dynamic RANS predictions and the experimental results is overall observed. The main difference with respect to the experimental realization consists in a reduced momentum defect in the proximity of the wall with respect to experimental data in the attached boundary layer under adverse pressure gradient. By recalling that the dynamic RANS solution employs a wall-modelled approach (the average thickness of the first layer of cells in the suction side is 50 friction units), these discrepancies with respect to experimental results are considered as acceptable. Regarding the standard $k-\omega$ SST formulation included in the comparison, the velocity profiles are reported only before separation. Indeed, as already pointed out in the previous section, the boundary layer predicted by the standard $k-\omega$ SST formulation does not separate -- it remains attached over almost the entire suction side of the wing ($x_s/c \approx 0.972$). As shown in figure \ref{fig:Velocity_profile_airfoil}, the standard $k-\omega$ SST formulation is found to overpredict the momentum defect, contrary to the dynamic RANS predictions.

\section{Conclusions}\label{sec_Conclusions}

Most of the theories on turbulence are based on a decomposition of the flow solution where the large coherent and anisotropic motions are separated from the small irregular and isotropic ones. The aim is to make the description and the understanding of turbulence more simple than that provided by the solution of the full Navier-Stokes equations. The most relevant example is provided by the groundbreaking theory for the small turbulent scales of Kolmogorov \citep{K41a, K41c}. Of equivalent relevance, the decomposition of turbulence is the fundamental basis for the development of coarse-grained approaches that represent the key enabling formalism for the establishment of a numerical solution of Navier-Stokes equations \citep{Germano1992}. In this respect, two main coarse-grained approaches have been established in the past that represent the theoretical basis for most turbulence closures, the Reynolds average and the spatial filtering formalism. The former has been widely used to develop statistical closures commonly called RANS models while the latter has been used to develop scale-resolving closures commonly referred to as LES models. However, coarse-grained approaches to turbulence have a degree of generality that goes beyond the Reynolds average and spatial filtering formalisms, as demonstrated by the fact that hybrid approaches simultaneously sharing statistical and scale-resolving features in different flow regions are often successfully used. In this context, a sufficiently general theoretical formalism able to describe coarse-grained approaches ranging from the statistical, hybrid and scale-resolving ones is provided by the temporally filtered Navier-Stokes equations. As demonstrated in \citet{Noi}, this alternative framework successfully describes coarse-grained approaches to turbulence and enables exact equations relating the turbulent stresses at different levels of coarsening. The latter have been used in \citet{Noi} to develop a new class of closures able to address both the statistical and scale-resolving approaches to turbulence. This class of closures is based on a dynamic computation of the model coefficient $c_\mu$ appearing in the definition of the eddy viscosity. In \citet{Noi}, the dynamic procedure has been developed for the $k-\varepsilon$ model and tested thus showing promising capabilities in accurately capturing statistical and scale-resolving features of turbulence. Among others, it is found that the dynamic $k-\varepsilon$ model is able to capture challenging flow features such as the transition to turbulence and the dependence of the reattachment length of separated boundary layers on the free-stream turbulence conditions.

In the present work, we further develop this class of turbulence closures by deriving two alternative formulations based on the $k-\omega$ model. The distinctive aspect of the two formulations is given by the possibility of using the dynamic coefficient in the destruction terms of the transport equations for $k_\tau$ and $\omega_\tau$. When this procedure is adopted, an equilibrium between the production and destruction terms of $k_\tau$ and $\omega_\tau$ is restored, thus leading to interesting flow features especially in aerodynamic applications. In these applications, free-stream turbulence plays a fundamental role in settling the solution around bodies especially in the cases of flow separations. The common practice is to model free-stream turbulence through the values of $k_\tau$ and $\omega_\tau$ imposed at the boundaries. However, these values decay from the inlet conditions thus not allowing us to model a constant ambient turbulence level surrounding the body. In this context, the adoption of the dynamic coefficient also in the destruction terms of the transport equations for $k_\tau$ and $\omega_\tau$ completely changes the scenario and allows us to have a zero decay of the transported turbulent quantities. This feature is claimed to be relevant especially when addressing the aerodynamic of elongated bodies.

Both the dynamic $k-\omega$ model and the zero-decay dynamic $k-\omega$ model have been tested in the transitional flow over a rounded leading edge flat plate. This flow is selected to address the capability of the modelling approach in capturing challenging phenomena like transition to turbulence and the dependence of the separation bubble length on the free-stream turbulence conditions. We found that both these flow phenomena are nicely reproduced by the dynamic $k-\omega$ model with and without zero decay. A net improvement of the results is observed when comparing the solution of the dynamic $k-\omega$ model with respect to that provided by its baseline closure, the standard $k-\omega$. An improvement in the quality of the results is also observed when comparing the solution with standard scale-resolving approaches such as the hybrid $k-\omega$ SST DES model. Of relevance, the new dynamic RANS model is found to be significantly more robust to a coarsening of the temporal resolution with respect to the hybrid $k-\omega$ SST DES model. In particular, we found that the quality of the solution remains reliable at remarkably coarse discretization levels up to $CFL \approx 75$ contrary to the hybrid $k-\omega$ SST DES model that loses its scale-resolving capabilities already for $CFL \approx 15$.

The zero-decay dynamic $k-\omega$ model has been also tested in the flow over a NACA 4412 at near-stall conditions. This flow has been selected to address the behaviour of the model when coupled with wall-modelling approaches which are necessary when using very-coarse wall-normal resolutions, typical of industrial applications. An agreement with experimental data is observed. In particular, the dynamic RANS approach is found to work again in capturing transition to turbulence, boundary layer separation and recirculating flow unsteadiness even when coupled with wall functions.

Finally, the effect of the baseline closure on the developed new class of dynamic RANS closures has also been addressed. We found that the solutions from the dynamic $k-\varepsilon$ and the dynamic $k-\omega$ exhibit very similar behaviours both quantitatively and qualitatively. This observation suggests that the dynamic procedure for the computation of the model coefficient $c_\mu$ smooths out the differences between the baseline closures. In particular, the dynamic procedure is found to resolve the different shortcomings that classically affect the $k-\varepsilon$ and $k-\omega$ closures, thus making the flow solution almost independent of the choice of the baseline model.

\bmhead{Research funding}
The research of this manuscript did not receive any funding.

%%%%%%%%%%%%%%%%%%%%%%%%%%%%%%%%%%%%%%%%%%%%%%%%%%%%%%%%%%%%%%%%%%%%%%%%%%%%%%%

\newpage
\begin{appendices}

\section{Dynamic $k-\varepsilon$ model}\label{sec_App_ke_dynamic}

The transport equations for the dynamic $k-\varepsilon$ model are given by
\begin{equation}
	\frac{\partial k_{\tau}}{\partial t} + \frac{\partial k_{\tau}\langle u_j \rangle_{\tau}}{\partial x_j}
	= 2 \nu_t \langle S_{ij} \rangle_{\tau} \langle S_{ij} \rangle_{\tau} - \varepsilon_{\tau}
	+ \frac{\partial}{\partial x_j} \left[ \left( \nu + \frac{\nu_t}{\sigma_k} \right) \frac{\partial k_{\tau}}{\partial x_j} \right] 
\end{equation}
\begin{equation}
	\frac{\partial \varepsilon_{\tau}}{\partial t} + \frac{\partial \varepsilon_{\tau}\langle u_j \rangle_{\tau}}{\partial x_j}
	= \frac{\varepsilon_{\tau}}{k_{\tau}} \left( C_{\varepsilon 1} 2 \nu_t \langle S_{ij} \rangle_{\tau} \langle S_{ij} \rangle_{\tau} - C_{\varepsilon 2} f_\epsilon \varepsilon_{\tau} \right)
	+ \frac{\partial}{\partial x_j} \left[ \left( \nu + \frac{\nu_t}{\sigma_\varepsilon} \right) \frac{\partial \varepsilon_{\tau}}{\partial x_j} \right]
\end{equation}
where $f_\epsilon$ is the near-wall damping function applied to the $\varepsilon_{\tau}$ destruction term, and is defined as
\begin{equation}
	f_\epsilon = \min \left[ \left( 1 - 0.3 \exp \left( - \left(\frac{R_{\tau}}{6.5}\right)^2 \right) \right) \left( 1 - \exp \left( - \frac{y^*}{3.1} \right) \right)^2, \: 1.0 \right]
\end{equation}
with the turbulent Reynolds number $Re_t = k_{\tau}^2 / (\nu \varepsilon_{\tau})$ and the dimensionless wall distance $y^* = y (\nu \varepsilon_{\tau})^{1/4} / \nu$ \citep{abe1995new}. The constants of the model are \citep{Launder}
\begin{align*}
	&C_{\varepsilon 1} = 1.5, \quad C_{\varepsilon 2} = 1.9 \\
	&\sigma_k = 1.4, \quad \sigma_\varepsilon = 1.4
\end{align*}
The eddy-viscosity is defined as
\begin{equation}
	\nu_t = c_\mu \frac{k_{\tau}^2}{\varepsilon_{\tau}}
\end{equation}
where the dynamic coefficient $c_\mu$, bounded $0 \leq c_\mu \leq 0.2$, is evaluated using the following dynamic procedure
\begin{equation}
	c_\mu = \frac{\mathcal{M}_{ij} \mathcal{L}_{ij}}{\mathcal{M}_{ij} \mathcal{M}_{ij}}
\end{equation}
with
\begin{equation}
	\mathcal{L}_{ij} = \langle \langle u_i \rangle_{\tau} \langle u_j \rangle_{\tau} \rangle_T - \langle \langle u_i \rangle_{\tau} \rangle_T \langle \langle u_j \rangle_{\tau} \rangle_T
\end{equation}
\begin{equation}
	\mathcal{M}_{ij}^{k-\varepsilon} = 2 \left \langle \frac{k_{\tau}^2}{\varepsilon_{\tau}} \langle S_{ij} \rangle_{\tau} \right \rangle_T - 2 \frac{k_{T_\tau}^2}{\varepsilon_{T_\tau}} \langle \langle S_{ij} \rangle_{\tau} \rangle_T
\end{equation}
The implementation of the model for the open-source software OpenFOAM \citep{OF} is available at the following link, \href{https://github.com/GiorgioManzi/DynamicEddyViscosity.git}{github.com/GiorgioManzi/DynamicEddyViscosity.git}.

\section{Zero-decay dynamic $k-\omega$ model}\label{sec_App_ko_zero_decay}

The modified transport equations for the dynamic $k-\omega$ model with the zero-decay adjustment are given by
\begin{equation}
	\frac{\partial k_{\tau}}{\partial t} + \frac{\partial k_{\tau}\langle u_j \rangle_{\tau}}{\partial x_j}
	= 2 \nu_t \langle S_{ij} \rangle_{\tau} \langle S_{ij} \rangle_{\tau} - c_\mu^* \beta^* k_{\tau} \omega_{\tau}
	+ \frac{\partial}{\partial x_j} \left[ \left( \nu + \frac{\nu_t}{\sigma_k} \right) \frac{\partial k_{\tau}}{\partial x_j} \right] 
\end{equation}
\begin{equation}
	\frac{\partial \omega_{\tau}}{\partial t} + \frac{\partial \omega_{\tau}\langle u_j \rangle_{\tau}}{\partial x_j}
	= 2 \gamma \frac{\omega_{\tau}}{k_{\tau}} \nu_t \langle S_{ij} \rangle_{\tau} \langle S_{ij} \rangle_{\tau} - c_\mu^* \beta \omega_{\tau}^2
	+ \frac{\partial}{\partial x_j} \left[ \left( \nu + \frac{\nu_t}{\sigma_\omega} \right) \frac{\partial \omega_{\tau}}{\partial x_j} \right]
\end{equation}
where $c_\mu^*$ is the dynamic coefficient $c_\mu$ normalised with respect to its conventional value $c_{\mu,0}=0.09$. The constants of the model are (\cite{Wilcox1988}):
\begin{align*}
	&\beta^* = 0.09, \quad \kappa = 0.41, \quad \sigma_k = 2.0, \quad \sigma_\omega = 2.0 \\
	&\beta = 0.075, \quad \gamma = \frac{\beta}{\beta^*} - \frac{\kappa^2}{\sigma_\omega \sqrt{\beta^*}}=0.52
\end{align*}
The eddy-viscosity is defined as
\begin{equation}
	\nu_t = c_\mu \frac{k_{\tau}}{\beta^* \omega_{\tau}}
\end{equation}
where the dynamic coefficient $c_\mu$, bounded $0 \leq c_\mu \leq 0.2$, is evaluated using the following dynamic procedure
\begin{equation}
	c_\mu = \frac{\mathcal{M}_{ij} \mathcal{L}_{ij}}{\mathcal{M}_{ij} \mathcal{M}_{ij}}
\end{equation}
with
\begin{equation}
	\mathcal{L}_{ij} = \langle \langle u_i \rangle_{\tau} \langle u_j \rangle_{\tau} \rangle_T - \langle \langle u_i \rangle_{\tau} \rangle_T \langle \langle u_j \rangle_{\tau} \rangle_T
\end{equation}
\begin{equation}
	\mathcal{M}_{ij}^{k-\omega} = 2 \left\langle \frac{k_{\tau}}{\beta^*\omega_{\tau}} \langle S_{ij} \rangle_{\tau} \right\rangle_T - 2 \frac{k_{T_\tau}}{\beta^*\omega_{T_\tau}} \langle \langle S_{ij} \rangle_{\tau} \rangle_T
\end{equation}
It should be noted that the standard dynamic $k-\omega$ formulation (which preserves free-stream decay) is perfectly analogous to the one presented above, with the sole exception being the absence of the $c_\mu^*$ coefficient in the destruction terms of both transport equations.

The implementation of the model for the open-source software OpenFOAM \citep{OF} is available at the following link, \href{https://github.com/GiorgioManzi/DynamicEddyViscosity.git}{github.com/GiorgioManzi/DynamicEddyViscosity.git}.

\end{appendices}

\bibliography{dynamic_rans_development.bib}% common bib file

\end{document}